%% file: H0_v5.5.tex
\documentclass[twocolumn,secnumarabic,amssymb, nobibnotes, aps, prd, aip, reprint]{revtex4-1}
\usepackage{graphicx}
\usepackage{dcolumn}
\usepackage{bbold}
\usepackage{amsmath}
\usepackage{amsfonts}
\usepackage{journal_names}
\usepackage{bm}
\usepackage{natbib}

\input{def2.tex}

\begin{document}

\title{The Hubble rate in averaged cosmology}

\author{Obinna Umeh, Julien Larena and Chris Clarkson}
\affiliation {Astrophysics, Cosmology and Gravity Center, and, Department of Mathematics and Applied Mathematics, University of Cape Town, Rondebosch 7701, South Africa}
\email{umeobinna@gmail.com,
julien.larena@gmail.com,
chris.clarkson@uct.ac.za}
\date{\today}

\begin{abstract}
The calculation of the averaged Hubble expansion rate in an averaged perturbed Friedmann-Lema\^itre-Robertson-Walker cosmology leads to small corrections to the background value of the expansion rate, which could be important for measuring the Hubble constant from local observations. It also predicts an intrinsic variance associated with the finite scale of any measurement of $H_{0}$, the Hubble rate today. Both the mean Hubble rate and its variance depend on both the definition of the Hubble rate and the spatial surface on which the average is performed. We quantitatively study different definitions of the averaged Hubble rate encountered in the literature by consistently calculating the backreaction effect at second order in perturbation theory, and compare the results. We employ for the first time a recently developed gauge-invariant definition of an averaged scalar. We also discuss the variance of the Hubble rate for the different definitions. 

\end{abstract}

\maketitle

\section{Introduction}
The late time Universe is not perfectly homogeneous and isotropic on all scales. The overdensities and voids that develop via gravitational collapse make it significantly inhomogeneous. 
As a result, the notion of a maximally symmetric background  geometry, which is  the very basic foundation  of the standard concordance model needs to be taken with extreme caution. Specifically, one would like to construct  an averaged model which can  suitably  describe the Universe on sufficiently large scales, as a coarse-grained version of the actual distribution of matter and energy in the Universe.
In the last decade, this issue has attracted considerable attention in cosmology, in particular through the so-called averaging problem (see \cite{Buchert:2007ik} and references therein).   This is significantly motivated by the belief that it could provide an answer to the Dark Energy problem  and the coincidence problem (see e.g. \cite{Rasanen:2003fy, Kolb:2004am, Rasanen:2006kp,Kolb:2005da}). Although it has not been shown to be the case,  the physics of averaging are still  worthy of  investigation because the parameters of cosmological concordance model are quite sensitive to the backreaction effect  \cite{Clarkson:2009hr} such sensitivity is very  important in the era of precision cosmology.

One method of  evaluating the backreaction effect lies within the standard cosmological model. That is, one can evaluate the backreaction from  perturbations of a background spacetime which describe structure formation.  At second-order, this effect gives rise to corrections to the local Hubble flow  \cite{Wang:1997tp,Clarkson:2009hr}. This idea was first investigated in an Einstein-de Sitter model in \cite{Kolb:2004am}, and followed up in more detail in \cite{Li:2007ci, Li:2007ny, Li:2008yj}.  This study was extended to include the case of a cosmological constant in \cite{Brown:2008ra, Brown:2009tg,Behrend:2007mf, Clarkson:2009hr}. However,  there appears to be some discrepancies between these results: While~\cite{Li:2007ci, Li:2007ny, Li:2008yj} found an important effect from backreaction, \cite{Brown:2008ra, Brown:2009tg,Behrend:2007mf, Clarkson:2009hr} found much smaller changes to the value of $H_0$ . Therefore our aim in this paper is to reconcile these results, and present them in a unified framework.  
  
We will make use of the averaging formalism  developed in \cite{Buchert:1999er, Larena:2009md, Clarkson:2009hr},  to estimate the corrections to  averaged local Hubble flow  by the small scale inhomogeneities in the matter distribution. 
Already  these  studies  \cite{Li:2007ci, Li:2007ny, Li:2008yj,Brown:2008ra, Brown:2009tg,Clarkson:2009hr},  have used this  formalism  to evaluate the corrections  to averaged  Hubble rate,  averaged deceleration parameter  and the equation of state of matter fluid,  to be expected from  the small scale inhomogeneities in the matter distribution, but in each case the definition of  averaged Hubble rate used is different. 
For instance, the authors of  \cite{Li:2007ci, Li:2007ny, Li:2008yj} ,  considered  a definition of  Hubble rate  which is very different  from the one considered in \cite{Brown:2008ra, Brown:2009tg} and also in \cite{Clarkson:2009hr}.  The authors in each case considered different slicing of the averaging domain, and/or different approaches to the cosmological  perturbation theory. Specifically in \cite{Brown:2008ra, Brown:2009tg} , the authors defined the averaged Hubble flow in the longitudinal gauge by following the expansion of the coordinate grids adapted to the gauge; this is an expansion rate which is  associated with the gravitational potential since the magnetic part of the Weyl tensor vanishes in this gauge~\cite{Clarkson:2009hr} (see also the Appendix of this paper for details on the magnetic part of the Weyl tensor).  This formalism  was later  used in \cite{Brown:2009tg} to study the effect of backreaction from averaging   in various gauges.
 
On the other hand, \cite{Li:2007ci, Li:2007ny, Li:2008yj} looked at the expansion of the matter fluid in the comoving synchronous gauge \cite{Li:2007ci, Li:2007ny, Li:2008yj}, while \cite{Clarkson:2009hr} studied the same thing, but  in longitudinal gauge.  Kolb et. al \cite{Kolb:2004am} studied the relationship between the averaged expansion of matter fluid in the comoving  synchronous gauge and the longitudinal gauge. The results and  claims  from these papers appear to differ base on the approach used.  It is against this backdrop that we propose to study  quantitatively  the difference between different definitions of Hubble rate.  This study will enable us to  clarify this issue and at the same time evaluate precisely the corrections to the concordance model as a result  of the backreaction effect.  We will also discuss how a consistent second order treatment of the backreaction effect in cosmological perturbation theory  changes  the value of the Hubble rate. The analysis of the intrinsic variance which  could affect the measurement of  the Hubble rate today, $H_{0}$, will be thoroughly investigated.

This paper is organized as follows: In Sections~\ref{sec1} and \ref{sec2}, we briefly recall the averaging formalism to be used, and discuss the definitions of the averaged Hubble rate in two different hypersurfaces of interest. In this section, we  will also use the gauge invariant formalism developed in  \cite{Gasperini:2009mu,Gasperini:2009wp} to calculate the averaged Hubble rate defined in the fluid frame. To the best of our knowledge, it is the first practical calculation making use of of this gauge invariant formalism to study backreaction effect. The  discussion of  our results is presented in Section~\ref{sec3} and  a fitting formula for the variance of the Hubble rate will be given here. We also  show that the two classes of definitions can be clearly  distinguished. A brief  comment on their relevance is also given. Finally, in Section~\ref{sec4}, we draw some conclusions and discuss future works. The Appendix presents the detailed expressions of the various Hubble rates considered in this paper, at second oder in cosmological perturbation theory.  

Throughout this paper, we will suppose that gravitation is well described by general relativity on all scales and that the cosmic matter fluid can be considered as a perfect fluid. Moreover, Latin letters of the beginning of the alphabet $(a,b,c,...,h)$ will denote spacetime indices, and Latin letters in the middle the alphabet $(i,j,k,...)$ will denote spatial indices.

\section{Equations of Motion}\label{sec1}

Buchert's averaging formalism ~\cite{Buchert:1999er,Buchert:2001sa} (a  similar  averaging formalism was presented earlier in \cite{Russ:1996km}) and its generalization to arbitrary coordinate systems~\cite{Larena:2009md, Brown:2009tg,Behrend:2007mf,Rasanen:2009uw} rely on Einstein equations written in the Arnowitt-Deser-Misner form. Within this formalism, one considers a set of observers defined at each point of the spacetime manifold, and characterized by a unit 4-velocity field, $n^{a}$,  that is everywhere timelike and future directed, i.e. $n^{a}n_{a}=-1$, with zero vorticity. This 4-velocity field induces a natural foliation of spacetime by a continuous set of space-like hypersurfaces locally orthogonal to $n^{a}$.  The projection tensor field onto these hypersurfaces is defined as  $h_{ab}=g_{ab}+n_{a}n_{b}$. 
The line element can then be written  with respect to this foliation:
\begin{equation}
\label{eq:ADM_metric}
ds^{2}=-(N^{2}-N_{i}N^{i}) d t^{2}+2N_{i} d t d x^{i}+h_{ij} d x^{i} d x^{j}\mbox{ ,}
\end{equation}
where we have introduced respectively the lapse function $N(x^{a})$ and the shift 3-vector $N^{i}(x^{a})$ . The components of the 4-velocity of the fluid comoving with the coordinate grids is given in relation with the lapse and shift  functions as
$n^{a}=\frac{1}{N}(1,-N^{i})\mbox{, } n_{a}=N(-1,0,0,0)\mbox{ .}
$\,
It is  orthogonal to the hypersurface $h_{ab}$ .
The intrinsic curvature of the hypersurfaces  is given by $\sr \equiv h^{ab}\sr_{ab}$, where $\sr_{ab}$ is the 3-Ricci curvature of the hypersurfaces and the extrinsic curvature (or second fundamental form): $K_{ab}\equiv -h^{c}_{a}h^{d}_{b}n_{c;d}$. 

Here we will consider only the Hamiltonian constraint  and the evolution equation for the metric of the spacelike hypersurface.(for the complete set of ADM decomposed Einstein equations see~\cite{Larena:2009md}). 
\begin{eqnarray}
 \left( \partial_t - \mathcal{L}_{\Sigma_t} \right) h_{ij} &=& - 2N K_{ij}   \label{Einstein_PDE1}\mbox{ ,} \\
R + K^2 - K_{ij} K^{ij} & =&16\pi \epsilon\mbox{ ,} \label{Einstein_PDE2}
 	\end{eqnarray}
	where
	 $\epsilon=n^a n^b T_{ab}$
	 and $T_{ab}$ is the energy momentum tensor defined to include the cosmological constant as
 $
T_{ab}=(\rho +p)u_{a}u_{b}+(p + \Lambda/(8 \pi G))g_{ab}\mbox{ .}
$
$u^a$ is time-like 4-velocity for the  matter field normalized to $u^a u_a=-1$, it is related to  $n^a$ through 
\begin{equation}
\label{eq:relvel}
u^{a}=\gamma (n^{a}+v^{a})\mbox{ ,}\mbox{ where } \gamma =\frac{1}{\sqrt{1-v^{a}v_{a}}}\mbox{ .}
\end{equation}
The vector $v^{a}$ is spacelike and  it is orthogonal to $n^{a}$ ($v^{a}n_{a}=0$).

The non-local, free gravitational field is described by the Weyl tensor. Given a timelike vector this is split into electric and magnetic parts. For example, with respect to $n^a$ these are
\be
E_{ab}^{(n)}=C_{acbd}n^cn^d~~~~\text{and}~~~~H_{ab}^{(n)}= {}^*\!C_{acbd}n^cn^d,
\ee
where $C_{abcd}$ is the Weyl tensor and $^*\!C_{abcd}$ is its dual. Analogous definitions exist for the vector field, $u^a$. This means that observers in the frame of the fluid and observers in the coordinate frame observe this electric-magnetic split differently (see~\cite{Maartens:1998xg} for the transformation relations between the two), analogously to boosted observers measuring different electric and magnetic parts of the electromagnetic field. In particular, in certain gravitational fields there may exist a special frame whereby one of these two components vanishes. For example, in so-called silent universes which are not conformally flat, there exists a preferred frame in which the magnetic part of the Weyl tensor is zero~-- such a frame may be considered the rest-frame of the gravitational field. In spacetimes where this is possible, it is unique as follows from the transformation laws in~\cite{Maartens:1998xg}, and there exist (at least) two physical, well motivated, frames: the rest-frame of the fluid, and the rest-frame of the non-local gravitational field.

So far we have defined two different 4-velocities, which according to standard $1+3$ decomposition of a covariant derivative of 4-vector, will imply defining respectively two expansion rates.

\subsection{Decomposition of velocities}\label{subsec1}
The covariant derivatives of the two $4-$velocities, $u^a$ and $n^a$, as well as the spacelike relative velocity $v^a$, may be invariantly decomposed with respect to the coordinate frame, $n^a$, (this corrects the  decomposition presented in  \cite{Larena:2009md}; however the expression for the Hubble rate is not affected):
\begin{eqnarray}
\nabla_{a}n_{b}&=&-n_{a}\dot{n}_{b}+\frac{1}{3}\xi h_{ab}+\Sigma_{ab}\mbox{ ,}\\
\nabla_{a}u_{b}&=&-\gamma v^c\left(\gamma^2\dot v_c+\dot n_c\right)n_an_b\nonumber\\
&&-\gamma \left(\gamma^2 v^c\tilde\nabla_a v_c+\frac{1}{3}\xi v_a+\Sigma_{ac}v^c\right) n_b\nonumber\\ &&
+\gamma n_a\left(\gamma^2 v^c\dot v_c v_b+\dot n_{\langle b\rangle}+\dot v_{\langle b\rangle}\right)\nonumber\\
&&+\frac{1}{3}\theta h_{ab}+\sigma_{ab}+\omega_{ab}\mbox{ ,}
%
\end{eqnarray}
\begin{eqnarray}
\nabla_{a}v_{b}&=&-\dot n_c  v^c\, n_a n_b - n_a\dot v_{\langle b\rangle}
+\left( \frac{1}{3}\xi v_a+\Sigma_{ac}v^c\right)n_b
\nonumber\\&&+\frac{1}{3}\kappa h_{ab}+\beta_{ab}+W_{ab}\mbox{ ,}\nonumber\\
\end{eqnarray}
where:
\begin{eqnarray}
 \xi\equiv h^{ab}\nabla_{a} n_{b}\mbox{ ,} \,\,\,\,\,\, \,\,\,\,\,\Sigma_{ab}\equiv h^{c}_{a}h^{d}_{b}\nabla_{(c}n_{d)}-\frac{1}{3}\xi h_{ab}\mbox{ ,} \,\,\,\,\, \,\,\,\,\nonumber\\ 
\theta\equiv h^{ab}\nabla_{a} u_{b} \mbox{ , } \,\,\,\,\,\,\,\,\,\,\,\sigma_{ab}\equiv h^{c}_{a}h^{d}_{b}\nabla_{(c}u_{d)}-\frac{1}{3}\theta h_{ab}\mbox{ ,} \,\,\,\,\,\,\,\nonumber\\ 
\omega_{ab}\equiv h^{c}_{a}h^{d}_{b}\nabla_{[c}u_{d]}\mbox{ , }\,\,\,\,\,\,\,\,\,\,\,\,\,\,\,\,\,
 \kappa\equiv h^{ab}\nabla_{a} v_{b} \mbox{ , }\,\,\,\,\,\,\,\,\,\,\,\,\,\,\,\,\,\,\,\,\,\,\,\,\nonumber\\
\beta_{ab}\equiv h^{c}_{a}h^{d}_{b}\nabla_{(c}v_{d)}-\frac{1}{3}\kappa h_{ab}\mbox{ , }\,\,\,\,\,\,\,\,
W_{ab}\equiv h^{c}_{a}h^{d}_{b}\nabla_{[c}v_{d]}\nonumber\mbox{ .} 
\end{eqnarray}
Where we have used the notation $\dot A_{a\cdots b}=n^c\nabla_c A_{a\cdots b}$, the  angle brackets denote symmetric, trace free, and projected with respect to $n_a$. $\tilde \nabla$ denotes the spatially projected covariant  derivative. 
 
Here $\xi$ and  $\theta$  are the expansion rates, while $\kappa$ is the divergence of the 3-velocity $v^a$; $\Sigma_{ab}$, $\sigma_{ab}$ and $\beta_{ab}$  are the shear, while $\omega_{ab}$ and $W_{ab}$ are the vorticity in the respective definitions.  Every quantity  defined here has a natural interpretation in terms of observers comoving with the fundamental $4-$velocity $n^a$.  Provided  these definitions are unique and consistent, all related quantities have a direct physical or geometric meaning with respect to the fundamental $4-$velocity $n^a$.   Any difference  between such 4-velocities will be of $\mathcal{O}(\epsilon)$ in perturbed FLRW case and will disappear in  the FLRW limit \cite{Maartens:1998xg} . Note that:
\be
\theta-\gamma\xi=\gamma\kappa\left(1+\frac{1}{3}\gamma^2v^2 \right)+\gamma^3v^av^b\beta_{ab},
\ee
in   FLRW limit,   $\gamma $ and  $v^2$ are of the order $\mathcal{O}(\epsilon^2)$, hence $\theta\sim\xi$. 
The decomposition of the matter 4-velocity, $u^a$,  is quite unusual, since it is with respect to $n^a$. One can calculate directly the normal acceleration, vorticity and shear and so on; for us the intrinsic expansion rate is important:
\bea
\Theta&=&\nabla_a u^a\nonumber\\ &=&
\theta+\gamma v^a\left(\gamma^2\dot v_a+\dot n_a \right).
\eea 

The following relations between expansion rates will be used later:
\begin{eqnarray}
\xi=\gamma^{-1}(\theta+\theta_{B})\mbox{ ,}\\
\Sigma_{ij}=\gamma^{-1}(\sigma_{ij}+\sigma_{Bij})\mbox{ .}
\end{eqnarray}
where $\theta_{B}\equiv -\gamma\kappa-\gamma^{3} B$ and for the shear:\\
$\sigma_{Bij}\equiv -\gamma\beta_{ij}-\gamma^{3}\left(B_{(ij)}-\frac{1}{3}Bh_{ij}\right)$.
The tensor $B_{ab}$ is defined as
\begin{eqnarray} 
B_{ab}&\equiv& \frac{1}{3}\kappa(v_{a}n_{b}+v_{a}v_{b})+\beta_{ca}v^{c}n_{b}\nonumber \\&& 
+\beta_{ca}v^{c}v_{b}+W_{ca}v^{c}n_{b}
+W_{ca}v^{c}v_{b}\mbox{ ,}
\end{eqnarray}
and its trace is given by $B=\frac{1}{3}\kappa v^{2}+\beta_{ab}v^{a}v^{b}$.

\section{Averaged Hubble rates}\label{sec2}
In general, the average of a scalar quantity  $S(t,x)$ may be defined as: 
\begin{equation}
\label{eq:Average}
 \average{S(t,x)}\equiv \frac{\int d^{3}x JS(t,x)}{\int d^{3}x J},
\end{equation}
where $J= \sqrt{h}$  is the  square root of the  determinant of the metric on the hypersurface orthogonal to $n^a$. 
The time derivative of Eq.~(\ref{eq:Average}) leads to a commutation relation \cite{Larena:2009md}
\begin{equation}
\label{eq:CommRel}
[\partial_{t}\cdot,\average{\cdot}]S(t,x^{i})=\average{N\xi S}-\average{N\xi}\average{S}\mbox{ ,}
\end{equation}
as is usual in the averaging context. 

There are different definitions of the  averaged Hubble parameter $H_\mathcal{D}$ in the literature, and we would like to be able to compare them in the context of the standard model, up to second-order in cosmological perturbation theory. We shall employ the longitudinal gauge below in order to calculate averages in the concordance model, which fixes our coordinate frame $n^a$. In the longitudinal gauge the magnetic part of the Weyl tensor vanishes, and the electric part is a pure potential field in the absence of anisotropic stress~\cite{Clarkson:2009hr} (see also Appendix \ref{appendA}), making this the rest-frame of the gravitational field, or Newtonian frame. In this sense, both $n^a$ and $u^a$ are physically well defined reference frames.

There are different local expansion rates:
\begin{itemize}
\item $\xi$: the expansion of the family of  coordinate observers. In the longitudinal gauge we employ below, this is the rest-frame of the gravitational field.
\item $\Theta$: The expansion of the fluid, as observed in the fluid rest-frame.
\item $\theta$: The expansion of the fluid, as observed in the gravitational rest-frame.  
\end{itemize}
When performing averaging, there are two spatial hypersurfaces of interest:
\begin{itemize}
\item $\average{~}$: Averaging in the gravitational frame.
\item $\langle~\rangle_{\mathcal{F}}$: Averaging in the rest frame of the fluid. 
\end{itemize}
Finally, when averaging expansion rates associated with the gravitational field,  there is the issue of the time coordinate to use: we can associate the time coordinate $t$ with the proper time of the `averaged observers', which, when using $n^a$ requires an extra factor of $N$ in the expansion rate~\cite{Clarkson:2009hr}.

\noindent {\it\textbf{Definitions based on $ \xi $}} 

As argued in ~\cite{Brown:2008ra,Gasperini:2009wp}, one can consider the evolution of the metric of the hypersurface: 
\begin{equation}\label{eq:metricevo}
 \partial_t h_{ij} = \frac{2}{3}N h_{ij} \xi + 2 N \Sigma_{ij} + D_iN_j+ D_jN_i
\end{equation}
and also assume that the dimensionless domain scale factor can be defined as 
$a_\mathcal{D}= \left(\frac{V_\mathcal{D}(t)}{V_\mathcal{D}(0)}\right)^{1/3} 
$
 where $V_\mathcal{D} $ is the volume of the domain. It is easy to show from  equation (\ref{eq:metricevo}),  that 
 \begin{eqnarray}
 3 H_\mathcal{D}&=& \frac{\partial_tV_\mathcal{D}}{V_\mathcal{D}}
=\frac{1}{V_\mathcal{D}}\int_\mathcal{D}\left( N \xi + D_k N^k\right) \sqrt{h}d^3x\nonumber\\
 &=&\average{N \xi + D_k N^k}\,.
\end{eqnarray}
This definition describes the average expansion  of the coordinate grid and says nothing directly about the matter field. It has been used in the recent literature for calculations in the longitudinal gauge~\cite{Brown:2009tg,Brown:2008ra}, in which case it can be interpreted as the expansion rate of the gravitational rest frame. We will find that this definition exhibits some interesting features, such  as  weak  scale dependence of backreaction effects.  

With this definition, the averaged Hamiltonian constraint ( \ref{Einstein_PDE2}) becomes:
\begin{eqnarray}
\label{HamilNxi}
 6H_{\CD}^{2}& =&16 \pi G \average{N^2 \gamma^2 \left( \rho + v^2 p\right) } + 2\average{N^2\Lambda} \nonumber\\ 
 & & - \average{N^2\mathcal{R}}- {\cal Q}_\CD +\mathcal{P}_\CD\mbox{ ,}\\
\mathcal{Q}_\CD&\equiv& \frac{2}{3}\average{N^2 \xi^2}-\frac{2}{3}\average{N \xi}^2 - 2\average{N^2 \Sigma^2} \mbox{ ,}\nonumber\\
 \mathcal{P}_\CD&\equiv& \frac{4}{3}\average{N \xi}\average{D_k N^k}+ \frac{2}{3}\average{D_k N^k}\mbox{ ,}\nonumber
 \end{eqnarray}
 where $\mathcal{Q}_\CD$ is the usual backreaction term and $ \mathcal{P}_\CD$ is an additional backreaction term which arises because of the inclusion of the shift parameter $N^k$.
 This  definition was used in \cite{Brown:2009tg,Brown:2008ra}.  Most studies have tended to include the lapse function in the definition. However, this choice is arbitrary (extensive discussion on this was given in \cite{Clarkson:2009hr}.).
One can also  choose to  define the Hubble factor without the lapse function as  
 $3 H_\mathcal{D}= \average{\xi}  
$ and the corresponding   averaged Hamiltonian constraint becomes
\begin{eqnarray}
\label{Hamilxi}
6H_{\CD}^{2} &=& 16 \pi G \average{ \gamma^2 \left( \rho + v^2 p\right) } + 2\average{\Lambda}  - \average{\mathcal{R }}\nonumber\\ 
 & & - {\cal Q_\CD}\mbox{,}\\ 
\mathcal{Q_\CD}&\equiv& \frac{2}{3}\average{ \xi^2}-\frac{2}{3}\average{ \xi}^2 - 2\average{\Sigma^2}\,.\nonumber
\end{eqnarray}

\noindent{\it\textbf{Definitions based on $\Theta$}}

Assuming all types of matter follow the same 4-velocity, the local expansion of the matter is given by $\Theta$. If we average this on spatial surfaces orthogonal to $u^a$, we have $3H_\mathcal{D}= \saverage{\Theta}_{\mathcal{F}}$. This definition is equivalent to that studied in \cite{Li:2007ny,Li:2008yj,Wang:1997tp}, and is the same as the expansion of the coordinates if we choose the synchronous gauge. The equations in that case are well known and presented in \cite{Buchert:2007ik}.  This choice might be most natural for  supernova observations, where the observer here on earth is  assumed comoving with the source.

\noindent {\it\textbf{Definitions based on $\theta$}}

A final definition of the expansion we consider is given by $\theta$: the derivative of the matter observers worldline projected into the rest-space of the gravitational frame.  This was introduced in~\cite{Larena:2009md,Clarkson:2009hr} as a way of recognising the fact the rest-frame of the matter before and after averaging are not the same. 
  Hence, a useful definition of the average Hubble factor is:
$3 H_\mathcal{D}=\average{N\theta}. 
$
This will lead to the following averaged Friedmann's equation:
\begin{eqnarray}
\label{HamilNtheta}
6H_{\CD}^{2}&=&16\pi G\left(\average{\gamma^{4}N^{2}\rho}+\average{\gamma^{2}(\gamma^{2}-1)N^{2}p}\right)\nonumber\\ &&
+2\Lambda\average{N^{2}\gamma^{2}}-\average{\gamma^{2}N^{2}\CR}-{\cal Q}_{\CD}+{\cal L}_{\CD}\mbox{ ,}\\
{\cal Q}_{\CD}&\equiv&\frac{2}{3}\left(\average{(N\theta)^{2}}-\average{N\theta}^{2}\right)-2\average{N^{2}\sigma^{2}}\mbox{ ,}\nonumber\\ 
{\cal L}_{\CD}&\equiv&2\average{N^{2}\sigma_{B}^{2}}-\frac{2}{3}\average{(N\theta_{B})^{2}}-\frac{4}{3}\average{N^{2}\theta\theta_{B}}\mbox{ .}\nonumber
\end{eqnarray}
where  $\sigma^{2}=\sigma^{i}_{j}\sigma^{j}_{i}/2$ and $\sigma_{B}^{2}=\sigma^{i}_{Bj}\sigma^{j}_{Bi}/2+\sigma_{ij}\sigma_{B}^{ij}$ had being adopted for simplification purposes.
In the same vein, we can also consider a definition of average Hubble factor without scaling  with lapse function as $3H_{\mathcal{D}}=\average{\theta }$, in this case the averaged Friedmann's equation becomes
\begin{eqnarray} 
\label{Hamiltheta}
6H_{\CD}^{2}&=&16\pi G\left(\average{\gamma^{4}\rho}+\average{\gamma^{2}(\gamma^{2}-1)p}\right)+2\Lambda\average{\gamma^{2}}\nonumber\\
 & &-\average{\gamma^{2}\CR}-{\cal Q}_{\CD}+{\cal L}_{\CD}\mbox{ ,}\\
{\cal Q}_{\CD}&\equiv&\frac{2}{3}\left(\average{(\theta)^{2}}-\average{\theta}^{2}\right)-2\average{\sigma^{2}}\mbox{ ,}\nonumber\\
 \mathcal{L}_{\CD}&\equiv& 2\average{\sigma_{B}^{2}}-\frac{2}{3}\average{(\theta_{B})^{2}}-\frac{4}{3}\average{\theta\theta_{B}}\mbox{ .}\nonumber
 \end{eqnarray}
 Notice that the Friedmann part of the  Buchert  equations averaged on the comoving hypersurface  may be recovered  from the last two definitions in the limit  where  $v_i\rightarrow 0$, $\gamma\rightarrow 1$ and $\average{\theta} \rightarrow\langle \Theta\rangle_{\mathcal{F}}$ \cite{Buchert:2007ik}.   
 The equations above [\ref{HamilNxi}-\ref{Hamiltheta}] contain the standard backreaction term ${\cal Q}_{\CD}$ and  additional backreaction term $\mathcal{L}_{\CD}$.  The additional  $\mathcal{L}_{\CD}$ term exist because of the non-vanishing peculiar velocity, $v^a$. In  almost FLRW metric, its contribution will be subdominant $\mathcal{O}(\epsilon^2)$. 

\subsection{Spatial averaging of a perturbed FLRW model}\label{subsec2}

The equations derived in Sec.~\ref{sec1} are not closed, but physical information can be extracted from them if we suppose that the Universe is well described by a perturbed FLRW  background. We shall consider perturbations in the longitudinal (Poisson) gauge, where the metric may be written as
\begin{equation}
\label{pertmetric}
ds^2=-\left(1+2\Phi+\Phi_2\right)dt^2+a^2\left(1-2\Phi-\Psi_2\right)\delta_{ij}dx^idx^j\mbox{ .}
\end{equation}
Here, the coordinates are chosen to coincide with the $n^a$ frame such that $n_a=-N\partial_a t$, 
where the lapse function is $
 N=\left(1+\Phi+\frac{1}{2}\Phi_2-\frac{1}{2}\Phi^2\right)$. We have used the trace-free part of the momentum constraint to set: $\Psi_1=\Phi_1=\Phi$ (that is, there is no anisotropic stress at first-order).  It was shown in \cite{Lu:2007cj,Lu:2008ju,Ananda:2006af,Matarrese:1997ay} that the vector and tensor modes induced by scalars  are subdominant  at second order, hence this line element is sufficiently accurate for our purposes. The peculiar velocity $v^i$ can be expanded to second order and is given by:
\[
 v_{i}=\frac{1}{2 a}\partial_{i}(2v_1+v_2).
\]
As usual, the background Friedmann's equation and the deceleration parameter  for the pure dust  and  positive cosmological constant universe are given by: 
\bea
H(z)^2&=&H_0^2\left[\Omega_0(1+z)^3+1-\Omega_0\right], \nonumber\\ 
q(z)&=&-\frac{1}{H^2}\frac{\ddot a}{a}=-1+\frac{1+z}{H(z)}\frac{d H}{d z}=-1+\frac{3}{2}\Omega_m(z)\nonumber
\eea
respectively, 
where 
\[
\Omega_m(z)=\frac{\Omega_0(1+z)^3}{\left[\Omega_0(1+z)^3+1-\Omega_0\right]^{1/2}}
\]
and the first order peculiar velocity given in terms $\Phi$ reads:
\be
v^{(1)}=-\frac{2}{3aH^2\Omega_m}\left(\dot\Phi+H\Phi\right).
\ee

For details about the solution to the first and the second order equations used in this work, see Appendix \ref{appendA}. 

In this framework, the average quantities on the hypersurface orthogonal to $n^a$  can easily be expanded to second order in perturbation theory, so that one would rather evaluate Euclidean integrals instead of a Riemann integrals~\cite{Kolb:2004am}:
\begin{equation}
\average{S}= S_0 +\saverage{S_1}+\saverage{S_2} +
\frac{1}{J_0}\saverage{S_{1}J_1}
-\frac{1}{J_0}\saverage{S_{1}}\saverage{J_1}.
\label{eq:AveFLRW}
\end{equation}
where $J_0$ and $J_1$ respectively stand for the background and the first order piece of  the square root of the  determinant of the $3-$metric, $\sqrt{h}$,  while $S_0 $, $S_1$ and $S_2$ are the background, first order and the second order component of any perturbed scalar on the hypersurface orthogonal to $n^a$. Note the important terms of the form $\saverage{~}\saverage{~}$ which appear due to the Riemann average~-- such terms do not appear if we average perturbations on the background only ( as in ~\cite{Baumann:2010tm} for example).

\subsubsection{Frame switching}\label{subsubsec2}

In order to perform spatial averages on the hypersurface comoving with the matter fluid, i.e. on the hypersurface orthogonal to $u^a$, while using the coordinate system of the longitudinal gauge presented in Eq.~(\ref{pertmetric}) we employ the technique developed in \cite{Gasperini:2009wp}. This will allow us to perform an average in a frame which is tilted with respect to the coordinates. We do this because the longitudinal gauge is well defined at second-order, and the solutions up to second-order are known in the case where the cosmological constant is non-zero~\cite{Bartolo:2005kv}. 

Before applying the formalism of \cite{Gasperini:2009wp} to the particular case of interest here, we summarise it and generalise it for our purposes. When defining the average of a spacetime scalar there is considerable freedom in the definition, and this freedom can be used to switch from an average defined in one frame to that in another (\cite{Gasperini:2009wp} used it to define gauge-invariant averages). Consider defining the average of a quantity using a spacetime window function $W_\Omega$:
\begin{equation}
 \langle S \rangle_{A_0,r_0} = {\int_{{\cal M}_4} d^4x \sqrt{-g} \, S \,W_\Omega(x) 
\over \int_{{\cal M}_4}  d^4x \sqrt{-g} \,W_\Omega(x)}\mbox{ ,}
\label{eq:Gasper4dint}
\end{equation}
where a a suitable  window function might be:
\begin{equation}
 W_\Omega(x)= \delta  (A(x)-A_0)H(r_0-B(x))\mbox{ .}
 \label{GasperWindow}
 \end{equation}
In this definition of the window function, $H$ is the Heaviside step function and $B(x)$ is a positive function of the coordinates with space-like gradient, $\nabla_a B$, and $A$ is a suitable scalar field with time-like gradient, $\nabla_a A$, such that it takes on a constant value $A_0$ on the hyper-surface of interest.  The scalar field $A$ then defines the foliation of spacetime for averaging.  The range of integration  across the  hyper-surface is specified by inserting a step-like definition of the spatial boundary  using the function $B(x)$, which is then bounded by a constant positive value $r_0 >0$.

It was argued in \cite{Gasperini:2009mu,Gasperini:2009wp} that one can consistently integrate out the coordinate time to define an average of the scalar field $S$ on the hypersurface of constant $A$ by performing a suitable change of coordinates that transforms the integration variable from $t\mapsto\tilde{ t}$. This can be achieved by defining $t= f(\tilde{ t}, x)$,  where the function $f$ is chosen to ensure that the scalar field $S$ transforms as $ S(f(\tilde{ t}, x),x) = \tilde{ S}(\tilde{t},x)$. By the use of the Jacobian factor $\partial t/\partial {\tilde{t}}$, the 3-metric is also transformed from  $h_{ij}$ into another metric $\tilde{h}_{ij}$. The function $f$ ensures that the scalar field $A(x,t)$ is homogeneous in the tilde frame: $A(f(\tilde{t},x),x)= \tilde A(\tilde{ t},x)\equiv A^{(0)}(\tilde{ t})$ (see \cite{Gasperini:2009wp} for details). Inserting this into Eq.~(\ref{eq:Gasper4dint}), one finds:
\begin{equation}
\label{37}
\langle S \rangle_{A_0}=
\frac{\int_{{\Sigma}_{A_0}} d^3x \tilde{J}\,~ \tilde{S}(t_0, {x})}
{ \int_{{\Sigma}_{A_0}}  d^3x \tilde{J}}\mbox{ ,}
\end{equation}
where the tilde quantities are evaluated in the new coordinate system.  According to \cite{Gasperini:2009wp}, this represents a  gauge invariant prescription for the average of a scalar object $S$ on the hypersurface $\Sigma_{A_0}$ of constant $A=A_{0}$. 

In cosmological perturbation theory, the square root of the  determinant of the metric $\tilde{g_{ab}}$ and the  scalar field $\tilde{S}$ can be expanded to second order in perturbation theory to give the average of a scalar field in the new coordinate system: 
\begin{equation}
\langle S \rangle_{A_0} = S_0 +\saverage{\tilde{ S_1}}+\saverage{\tilde{S_2}} +
\frac{1}{\tilde{J_0}}\saverage{\tilde{J_1} \tilde{S}_{1}}
-\frac{1}{\tilde{J_0}}\saverage{\tilde{S}_{1}}\saverage{\tilde{J_1}}\mbox{,}
\label{310}
\end{equation}
where $\tilde{J_0}= \sqrt{-\tilde{g}_{0}}$ and $\tilde{J_1}=\sqrt{-\tilde{g}_{1}}$ are  the background and the first order piece of  the square root of the determinant of the metric $\tilde{g}_{ab}$ respectively. The metric  $\tilde{g}$ is still a 4-dimensional metric.
By making a gauge transformation~\cite{Bruni:1996im} back to the original  coordinates, we obtain:
\begin{eqnarray}
\langle S \rangle_{A_0}&= &S_0 +\saverage{S_1}+\saverage{S_2}+\frac{1}{J_0}\saverage{J_1S_1}-\frac{1}{J_0}\saverage{S_1}\saverage{J_1}\nonumber\\
 & &-\frac{\dot S_0}{\dot A_0}\left[\saverage{A_1}+\frac{1}{J_0}\saverage{J_{1} A_1}+\saverage{A_2}\right]\nonumber\\
 & &
 +2\frac{\dot S_0}{\dot A^2_0}\saverage{A_1\dot A_1}-\frac{1}{\dot A_0}\left[\saverage{A_1\dot S_1}+\saverage{S_1\dot A_1}\right]\nonumber\\
 & &+\frac{1}{2}\left[\frac{\ddot S_0}{\dot A_0^2}-3 \frac{\ddot A_0\dot S_0}{\dot A^3_0}
 +2 \frac{\partial_t\left(\ln J_0 \right)\dot  S_0}{\dot A_0^2}\dot S_0\right]\saverage{A_1^2}\nonumber\\
 & &+\left[\frac{\ddot A_0}{\dot A_0^2}-\frac{\partial_t \left(\ln J_0\right)}{\dot A_0}\right]\saverage{S_1A_1}+ 2\frac{\dot S_0}{J_0\dot A_0}\saverage{A_1}\saverage{J_1}\nonumber\\  
 & &
 -\left[\frac{\ddot A_0}{\dot A_0^2}-\frac{\partial_t\left(\ln J_0\right)}{\dot A_0}\right]\saverage{S_1}\saverage{A_1}\nonumber\\ 
 & &  
-\left[\frac{\dot S_0\ddot A_0}{\dot A_0^3}+ \frac{\partial_t\left(\ln J_0\right)\dot S_0}{\dot A_0^2}\right]\saverage{A_1}^2\nonumber\\ 
 & &  
 -\frac{\dot S_0}{\dot A_0^2}\saverage{A_1}\saverage{\dot A_1}+\frac{1}{\dot A_0}\saverage{S_1}\saverage{\dot A_1}.
 \label{eq:Gasper}
\end{eqnarray}
Here $J_0= \sqrt{-g_{0}}$ and $J_1=\sqrt{-g_{1}}$ are  the background and the first order piece of the square root of the  determinant of the four dimensional  metric $g$  respectively. This is the major difference between this averaging prescription and the conventional one defined in equation (\ref{eq:AveFLRW}).
Once the scalar variable $A$ is chosen to specify the averaging hypersurface, the above averaging prescription can easily be applied. Eq.~(\ref{eq:Gasper}) was first derived in~\cite{Gasperini:2009wp}, but the authors set the spatial average of a first order scalar quantity $\saverage{S_1}$ to zero (see Eq.~(3.10) in \cite{Gasperini:2009wp}) before performing the gauge transformation, thereby neglecting the terms of the form $\saverage{S_1}\saverage{A_1}$, $\saverage{S_1}\saverage{\dot A_1}$, etc, which are non-zero and are explicitly scale dependent at second order~\cite{Clarkson:2009hr}. We have inserted them as they play an important role in the average of the Hubble rate.

To fix the definition of $A$ in terms of the quantities of the perturbed FLRW background and at the same time fix the foliation of interest, we employ the technique used in~\cite{Kolb:2004am}. This involves relating the scalar field $A$ to the time, $\tau$, measured by the observers with 4-velocity $u^a$ comoving with the fluid:
$u^0\partial_0 + u^i\partial_i= \partial_{\tau}.\,
$
 The scalar field $A$ can be expanded to second order in perturbation theory, subject to the condition $ \tilde A (t,x) = A_0(t) + A_1(t,x)+A_2(t,x)
\equiv\tau$ ~\cite{Gasperini:2009mu}
to give (using $u^a \nabla_a \tau=1$): 
\begin{eqnarray}
(1-\Phi-\frac{1}{2}\Phi_2+\frac{3}{2}\Phi^2+ v_1^k v_{1k})\partial_t \tilde A (t,x)\\\nonumber
+\frac{1}{a^2}(v_1^i+v_2^i)\partial_i \tilde A (t,x)=1\mbox{ .}
\end{eqnarray}
We can now calculate the higher order $A$ in terms of $\Phi_1$ and $\Phi_2$ of the perturbed FLRW background. This gives:
\bea
A_0(t)&=&t,\\
A_1(t,x)&=& \int_0^t \Phi_1 dt ,\\ 
A_2(t,x)&=& \frac{1}{2}\int_0^t \Phi_2 dt - \frac{1}{2} \int_0^t\Phi_1^2 dt - \int_0^t v_1^k v_{1k} dt\nonumber\\
 & & - \int_0^t \frac{1}{a^2}v_1^i\partial_i A_1 dt\mbox{ .}
\eea
The average Hubble factor calculated using this prescription is given in the Appendix B.

\subsection{The ensemble average and the variance}

With the tools developed in  Section. \ref{subsec2},  we have performed a consistent second order perturbative expansion of the Riemann average defined in Sec. \ref{subsec1} to obtain a corresponding Euclidean average. Given a specific realisation of a cosmology, we could now calculate spatial averages directly. Alternatively, we can calculate the ensemble average of a given spatial average which will tell us the expectation values of spatially averaged quantities. The ensemble-variance tell us how much we expect that to vary from one domain to another.

The ensemble average of a spatial average may be written as:
\begin{equation}
\overline{\saverage{X}}=\frac{1}{V}\int d^3x W(x/R_\mathcal{D})\overline{X(\bm x)},
\end{equation}
where the overbar denotes an ensemble average. We have specified the domain though the window function $W$. The Euclidean volume of the spatial domain of averaging $\CD$ is then given by:
$V=\int d^3 x W(x/R_\CD)$ which in the case of a Gaussian window function which we mostly employ is $V=4\pi R_\CD^3\int_0^\infty y^2 W(y) d y = (2\pi)^{3/2} R_\CD^3
$
for any $R_{\CD}$. The inverse Fourier transform of this window function reads:
$
W(kR_\CD)=\frac{1}{V}\int d^3x W(x/R_\CD) \, e^{-i\bm k\cdot \bm x}\mbox{ .}
$
The Fourier and the inverse Fourier transforms of any scalar quantity $\Phi$ are given as 
\begin{eqnarray}
 \Phi(\bm x)& = &\frac{1}{(2\pi)^{3/2}}\int d^3k\Phi(\bm k) e^{i \bm k \cdot\bm x}\mbox{,}\\ \Phi(\bm k)& =& \frac{1}{(2\pi)^{3/2}}\int d^3x\Phi(\bm x) e^ {-i \bm k \cdot \bm x}\mbox{.}
\end{eqnarray}
For statistically homogenous Gaussian random variables, we have:
$
\overline{\Phi(\bm k)}  =  0 \mbox{,} 
$
and the power spectrum of $\Phi$ is defined  by
\begin{equation}
\overline{\Phi(\bm k)\Phi(\bm k')}=\frac{2\pi^2}{k^3}\mathcal{P}_\Phi(k)\delta(\bm k+\bm k'). 
\end{equation}
Assuming scale-invariant initial conditions from inflation, this is given by
\be
\mathcal{P}_\Phi(z,k)=\left( \frac{3 \Delta_\mathcal{R}}{5
g_{\infty}} \right )^2 g(z)^2 T(k)^2  
\ee
where $T(k)$ is the normalised transfer function, $\Delta_{\mathcal{R}}^{2}$ is the primordial power of the
curvature perturbation~\cite{Komatsu:2008hk}, with
$\Delta_{\mathcal{R}}^{2} \approx 2.41 \times 10^{-9}$ at a scale
$k_{CMB}=0.002 \mathrm{Mpc}^{-1}$ (for the definitions of $g(z)$ and $ g_{\infty}$, please see equation \ref{gfac} in Appendix \ref{appendA}).

It is not difficult to notice from the equations displayed in the appendix that most of the terms we are dealing with are scalars which schematically appear in the form $ \partial^m\Phi( x)\partial^n\Phi( x)$ where $m$ and $n$ represent the number of derivatives (not indices), such that $m+n$ is even so that there are no free indices. (For example, $\partial_i\Phi \partial^2 \partial^i\Phi$ has $m=1$ and $n=3$.) Then from the results of \cite{Clarkson:2009hr}, the ensemble average of these kind of terms, if a Gaussian window function is assumed, is given by:
\begin{equation}
 \overline{\saverage{\partial^m\Phi( x)\partial^n\Phi( x)}}
=\frac{(-1)^{(m+3n)/2}}{2\pi^2}\int{dk} \, k^{m+n-1} k^3\mathcal{P}_\Phi(k).
\end{equation}

Using $\Phi = g(t) \Phi_0 (x)$, $g(t)$ being the growth suppression  factor and $\Phi_0(x)$ the   spatial dependent initial condition (see the Appendix), the terms that appear with a time derivative of the gravitational potential can be re-written to pull out the time component before evaluating the ensemble average:
\begin{equation}
 \dot\Phi(t, x)=-\left( 1+z\right) H \frac{d \ln g}{d z}\Phi(t, x).
\end{equation}
For the details of the calculation of the ensemble average of the inverse laplacian appearing the second order Bardeen potential refer to~\cite{Clarkson:2009hr}.

The ensemble variance in the Hubble factor is given by
\beq
\mbox{Var}[H_\mathcal{X}]=\overline{H_\mathcal{X}^{2}}-\overline{H}_\mathcal{X}^{2}\mbox{ ,}
\eeq
where $H_{\mathcal{X}}$ can be any definition of  averaged expansion rate we are studying. With this definition, it is easy to see that pure second order contributions drop out of the variance, so that only terms that are quadratic in first order quantities remain.

\section{Results and Discussion}
\label{sec3}

We shall now investigate the expectation values of the different average Hubble rates, along with their variances. For this we will consider an Einstein-de Sitter model, and a standard concordance model. 
We shall use length scales intrinsic to the model as reference points for  averaging:  these scales are the equality scale, $k_\text{eq}^{-1},$ and the Hubble scale, $k_\text{H}^{-1}$:
\begin{eqnarray}
k_\text{eq}&\approx& 7.46 \times 10^{-2}\Omega_0h^2\text{Mpc}^{-1},\\ \nonumber
k_\text{H}&\approx&\frac{h}{3000}\text{Mpc}^{-1},
\end{eqnarray}
where $\Omega_b$ and $\Omega_0$ are the baryon and total matter contributions today and $H_0=100 \,h\,$kms$^{-1}$Mpc$^{-1}$.
We shall also  use two models for comparison.  The first  is an Einstein-de Sitter model  with $h=0.7$ and 5\% baryon fraction (WMAP~\cite{Komatsu:2008hk} whose  estimate as an energy density is given as  $\Omega_b\approx0.046$). This has $k_ \text{eq}^{-1}\simeq 27.9$Mpc. The other model we shall use is the concordance model with $\Omega_0=0.26, h=0.7, f_\text{baryon}=0.175$ (this is the WMAP best fit~\cite{Dunkley:2008ie}). The key length scales in these model are $k_ \text{eq}^{-1}\simeq 107.2$Mpc and the Hubble scale $k_ \text{H}^{-1}\simeq4.3$Gpc. 

To calculate the integrals we shall use the  transfer functions presented in~\cite{Eisenstein:1997ik}. All lengths scales  shown are in Mpc unless otherwise stated. Because some of the integrals have a logarithmic IR divergence, all $k$-integrals have an IR cut-off set at ten times the Hubble scale, it did not appear explicitly in any of our calculations.   Moreover, the position of the  IR cut-off  does not affect the result, that is one can set the IR cut-off within this range $10-10^9$ and the results shown here will remain unchanged.  Since the divergence is logarithmic,  the result depends very mildly on where the cut-off is set.

\begin{figure}[htb!]
Einstein de Sitter\hfil LCDM\\
\includegraphics[width=0.4\columnwidth]{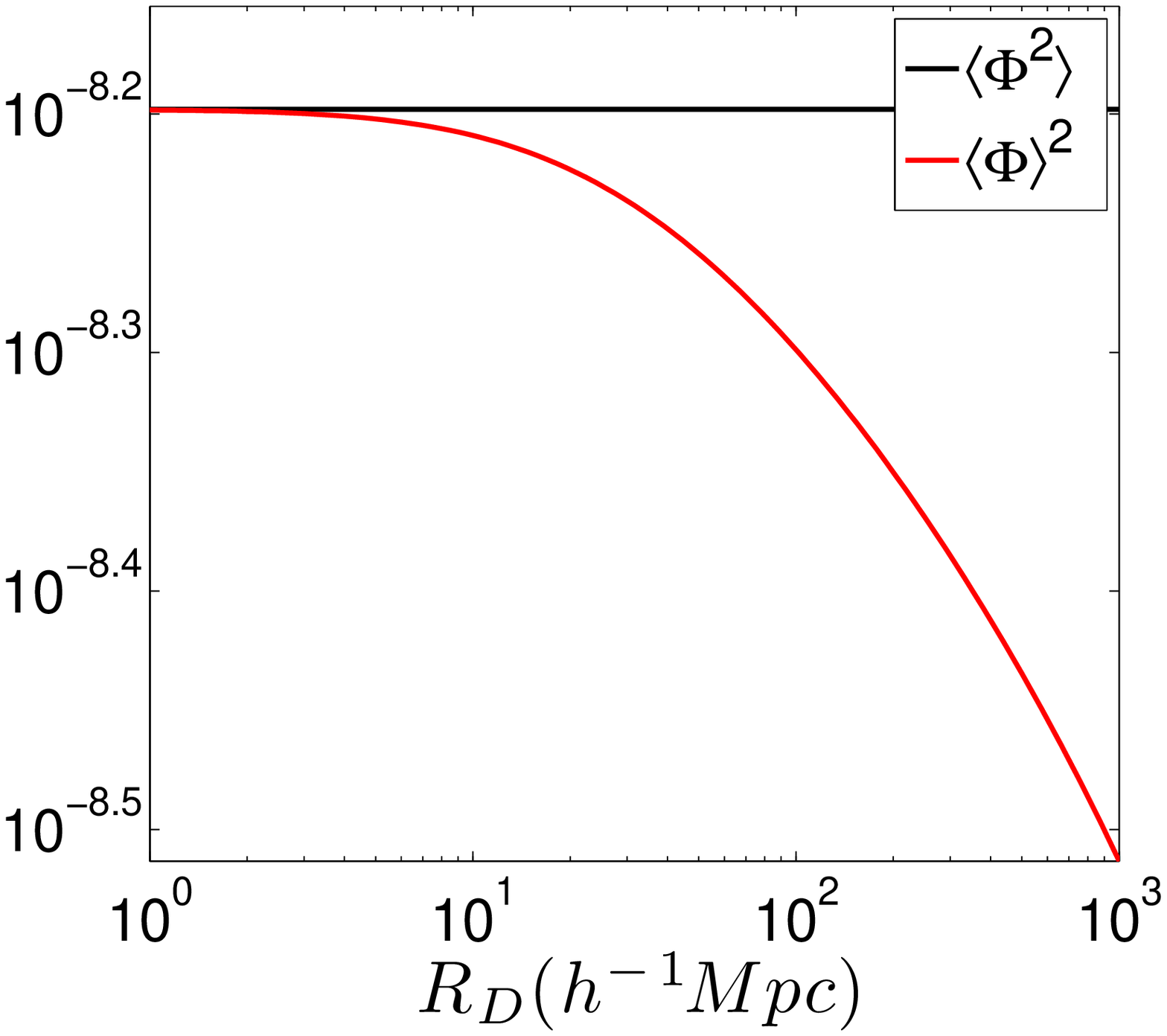} 
\includegraphics[width=0.4\columnwidth]{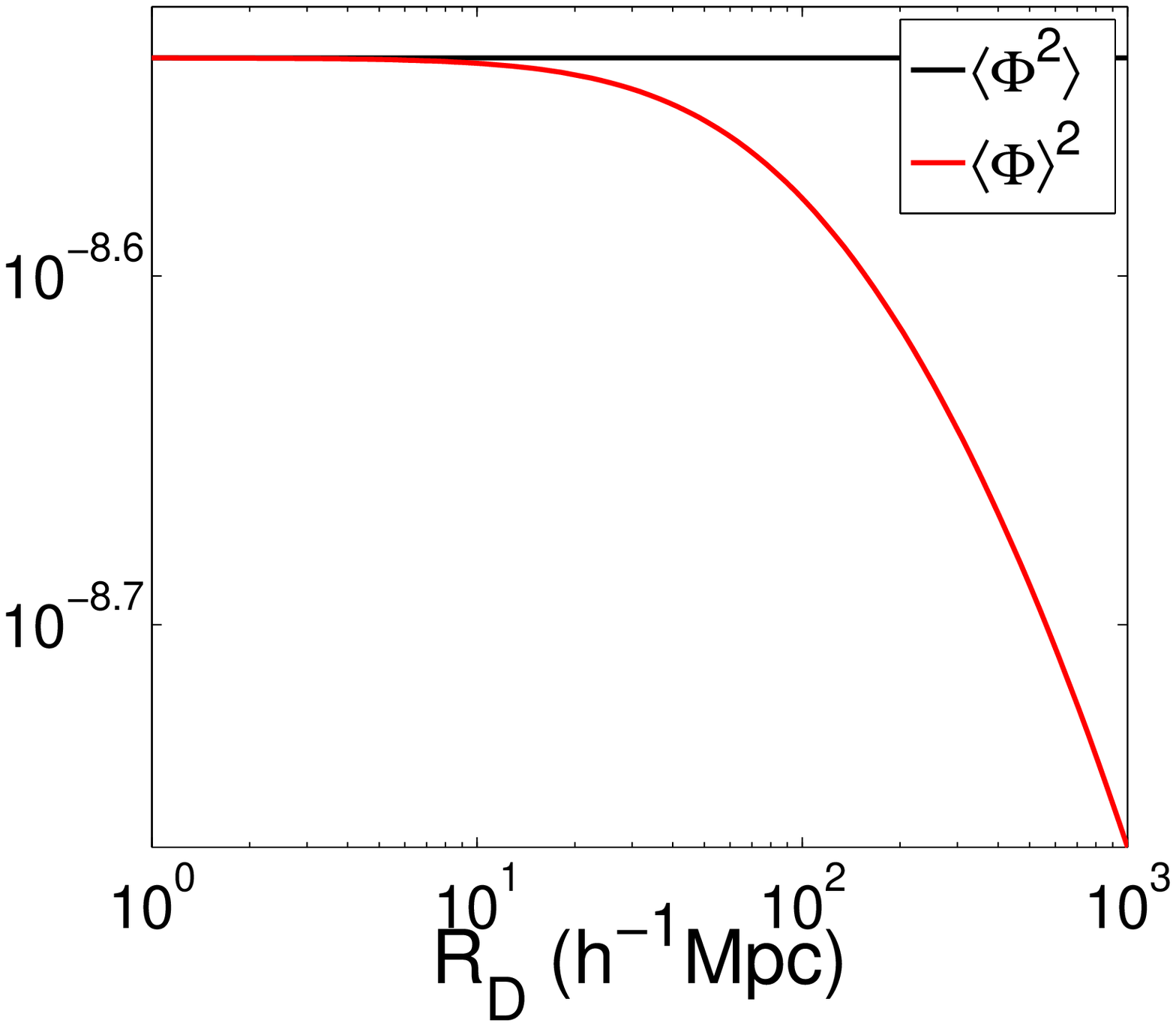}
\includegraphics[width=0.4\columnwidth]{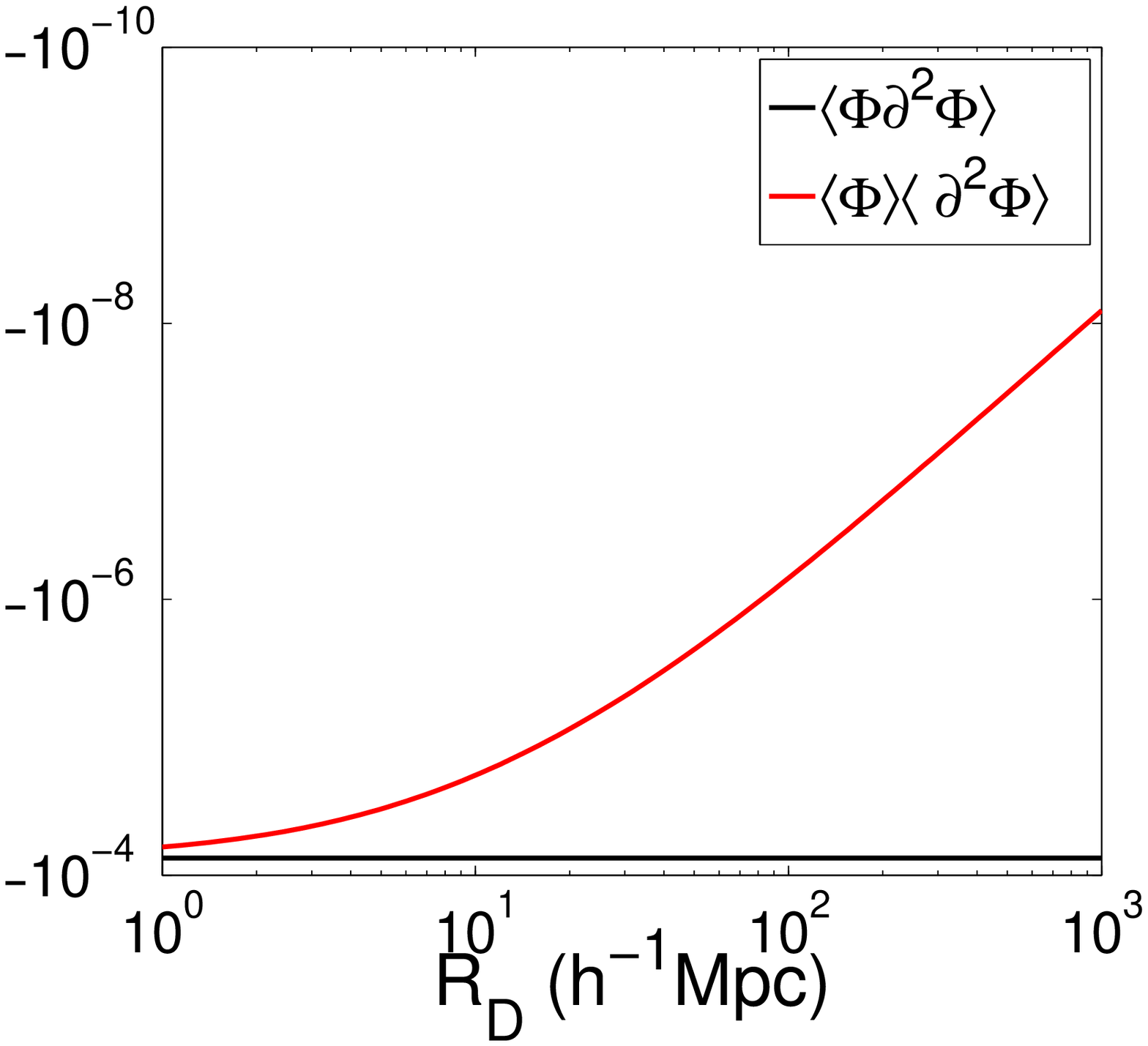}
\includegraphics[width=0.4\columnwidth]{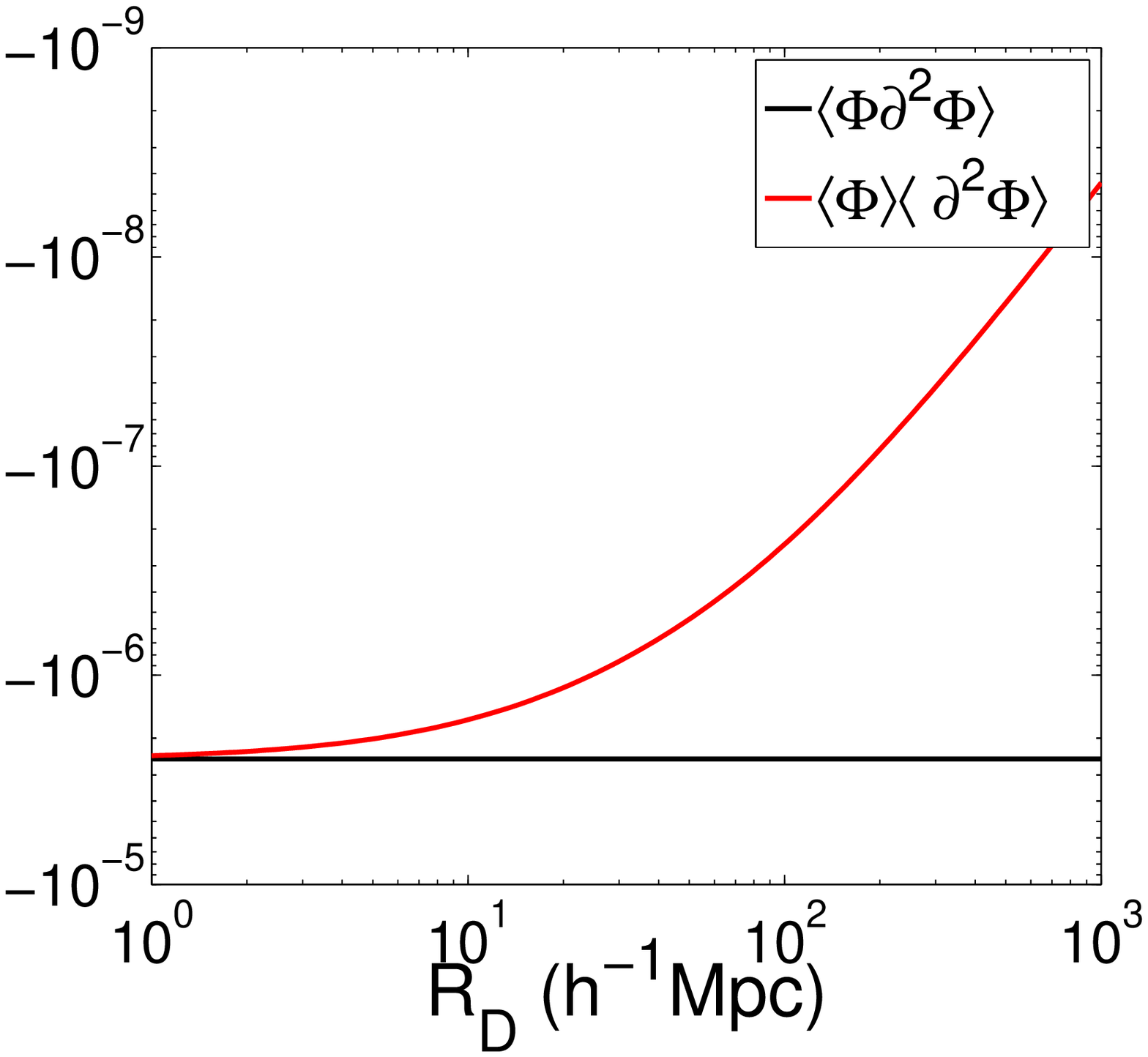}
\includegraphics[width=0.4\columnwidth]{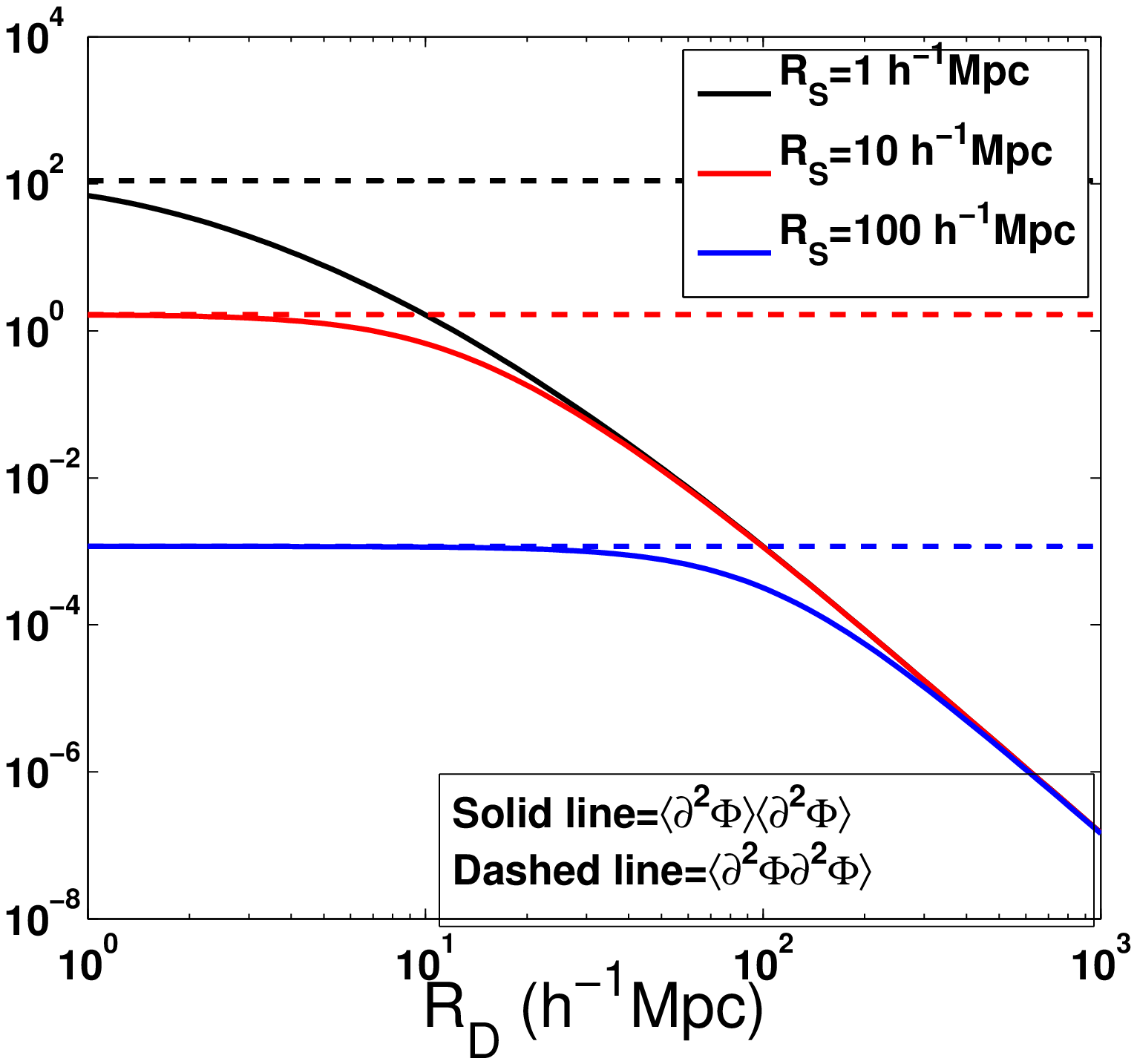}
\includegraphics[width=0.4\columnwidth]{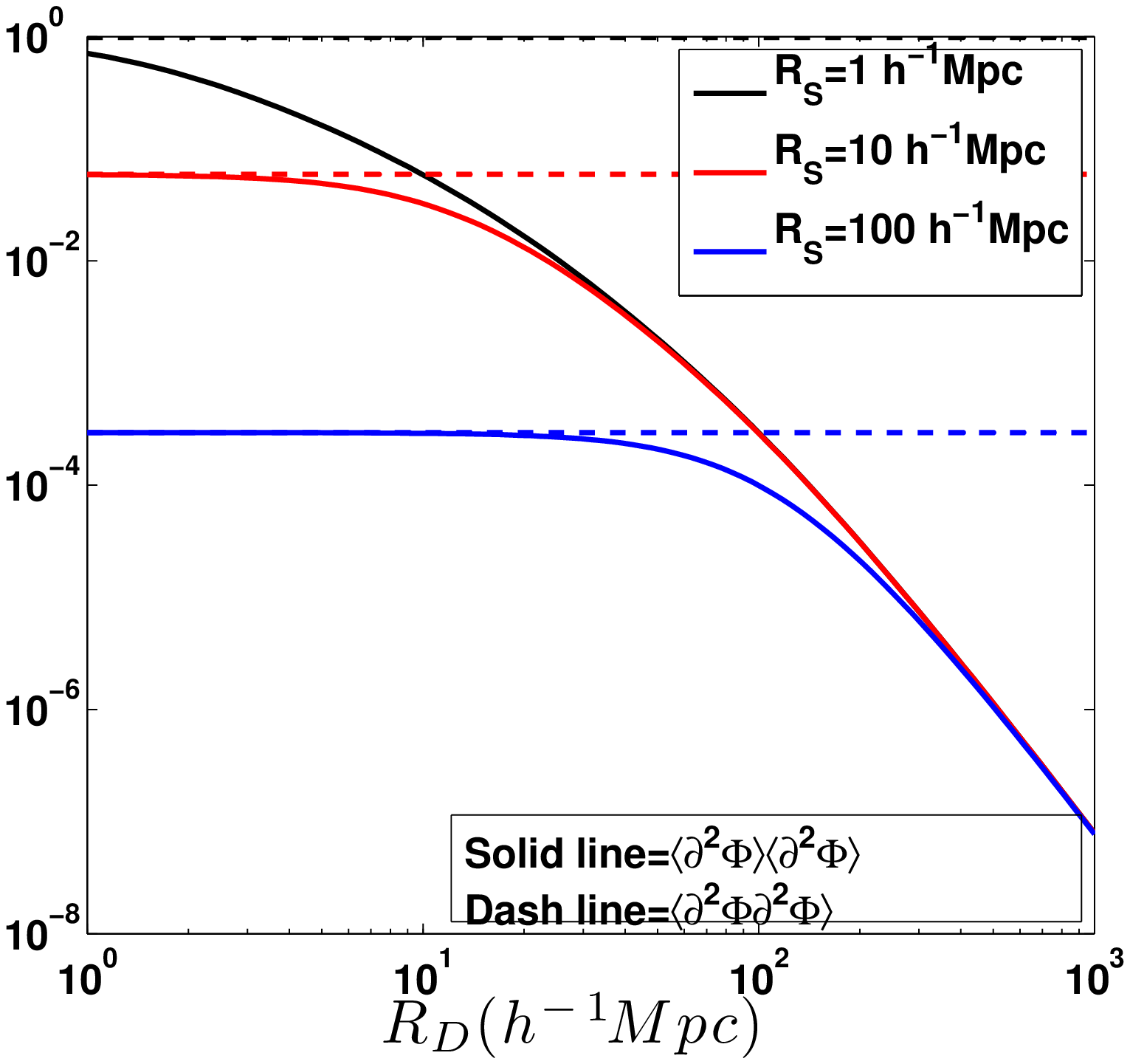}
\includegraphics[width=0.4\columnwidth]{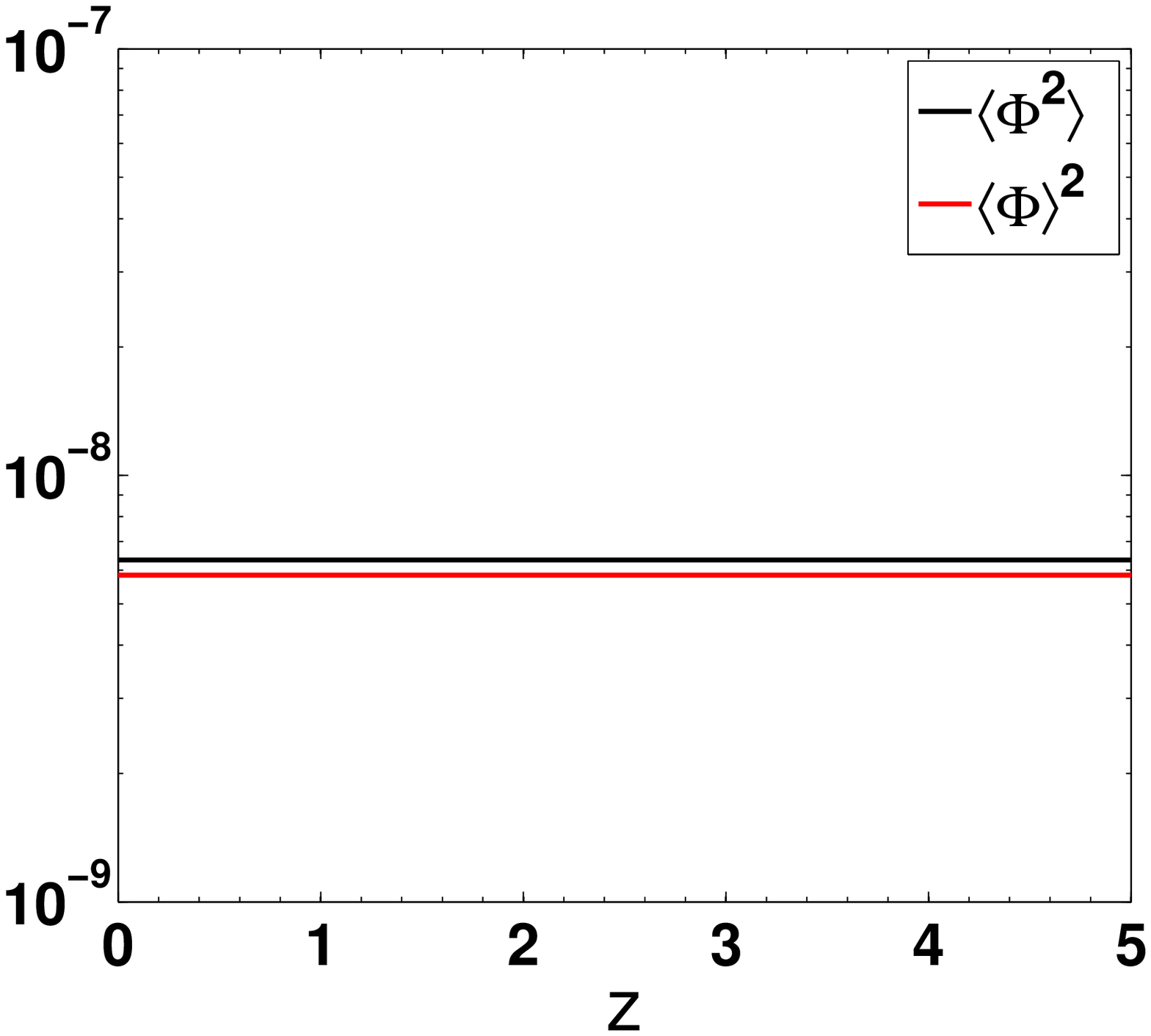}
\includegraphics[width=0.4\columnwidth]{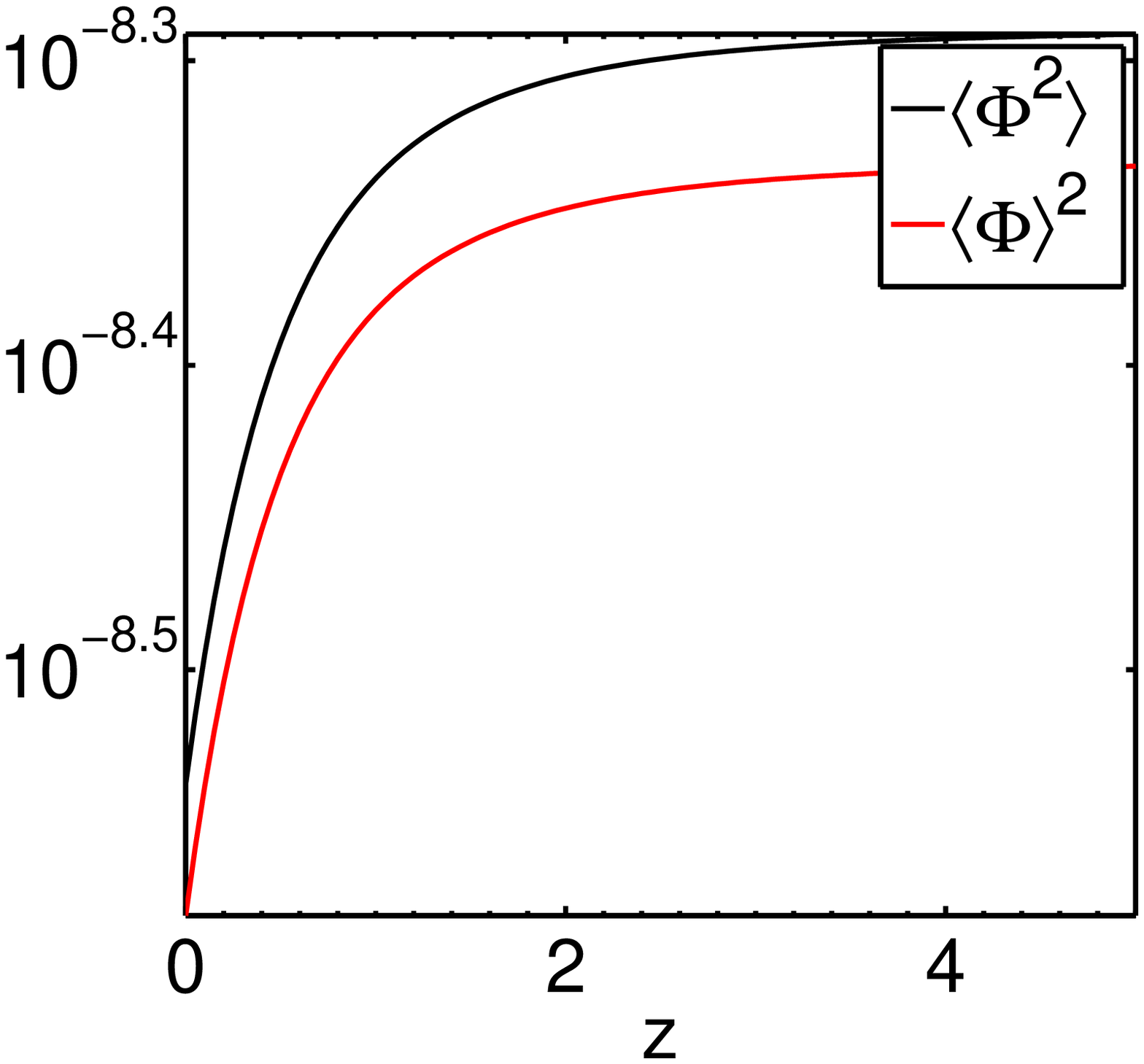}
\includegraphics[width=0.4\columnwidth]{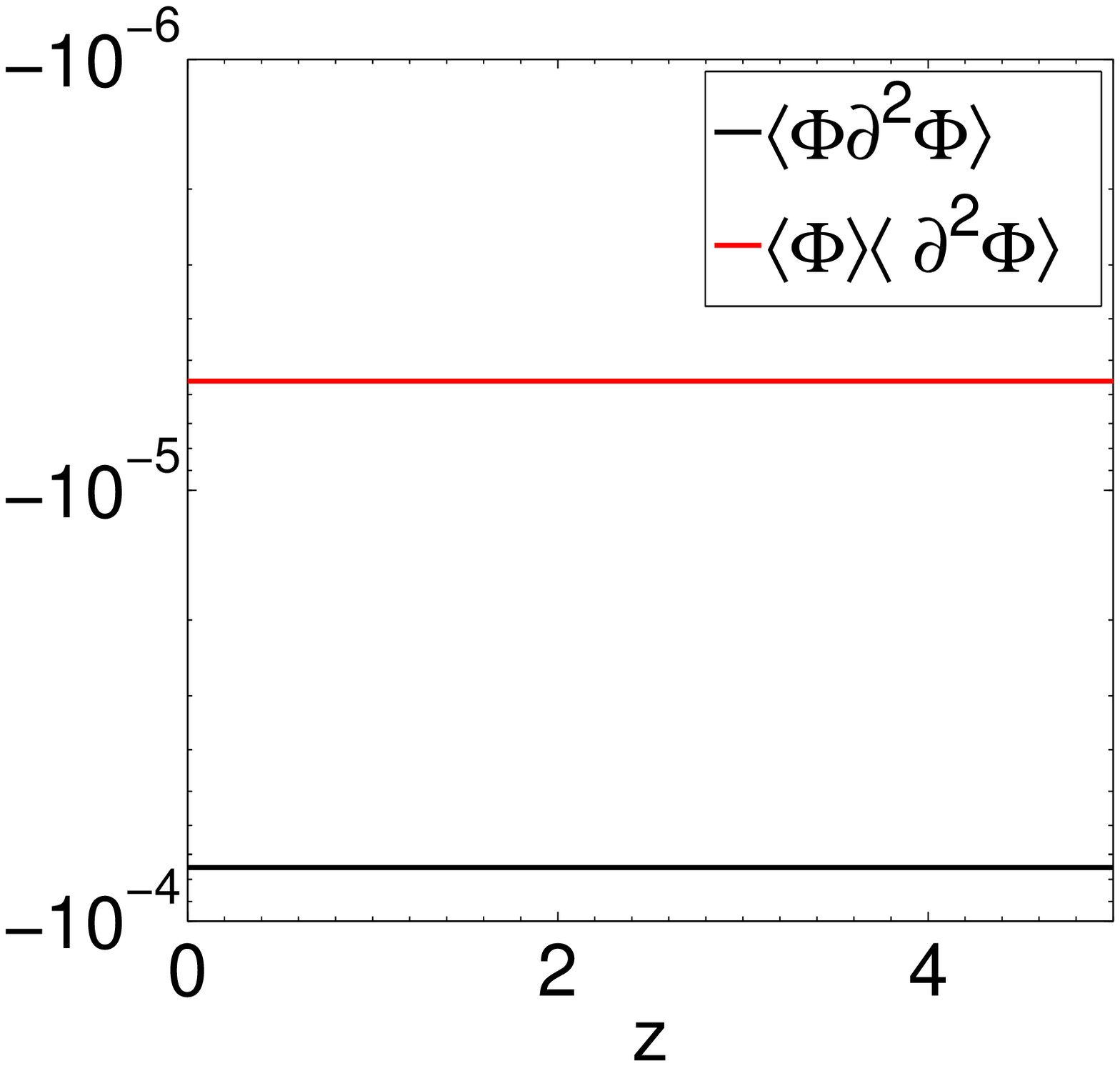}
\includegraphics[width=0.4\columnwidth]{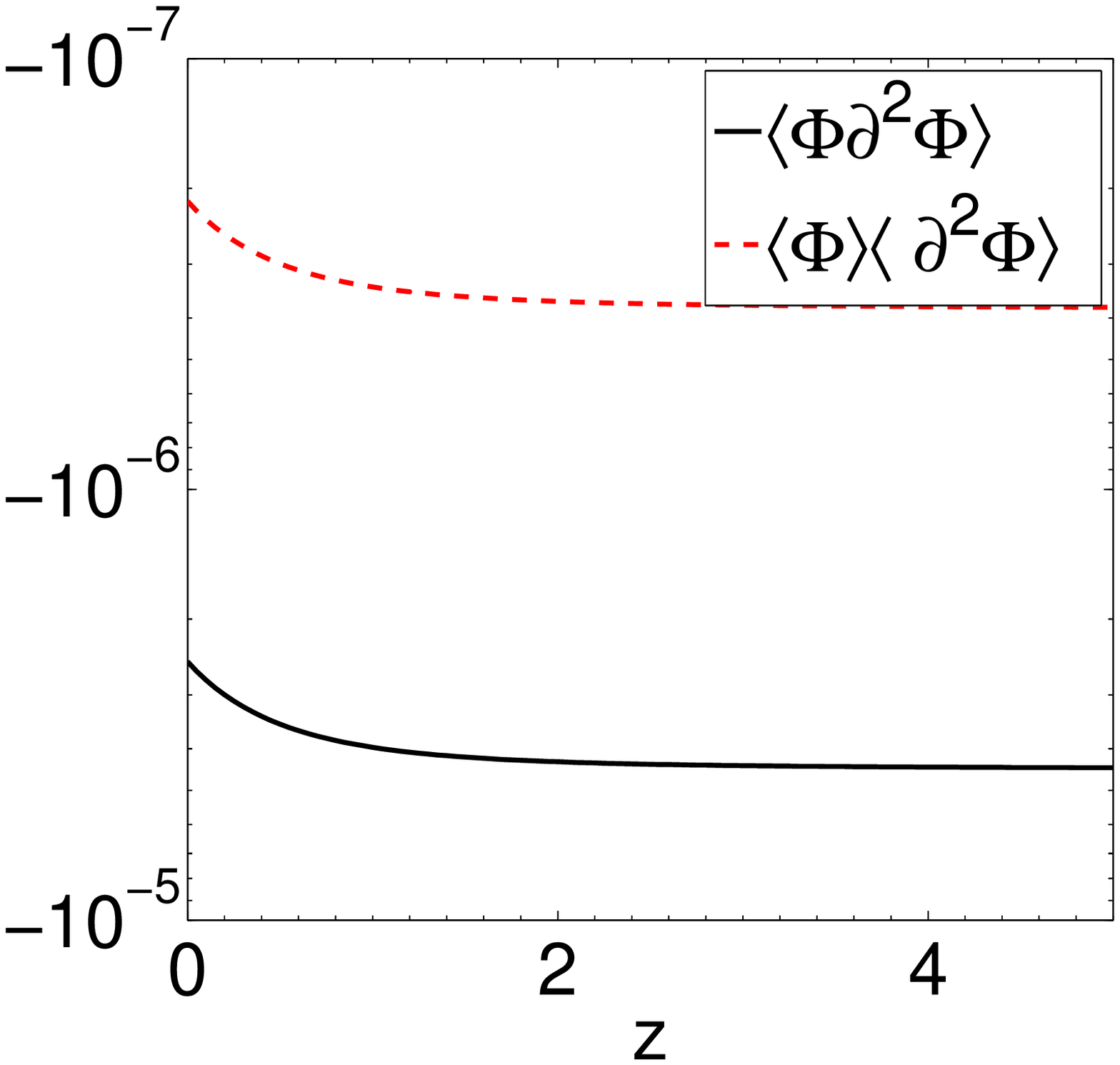}
\includegraphics[width=0.4\columnwidth]{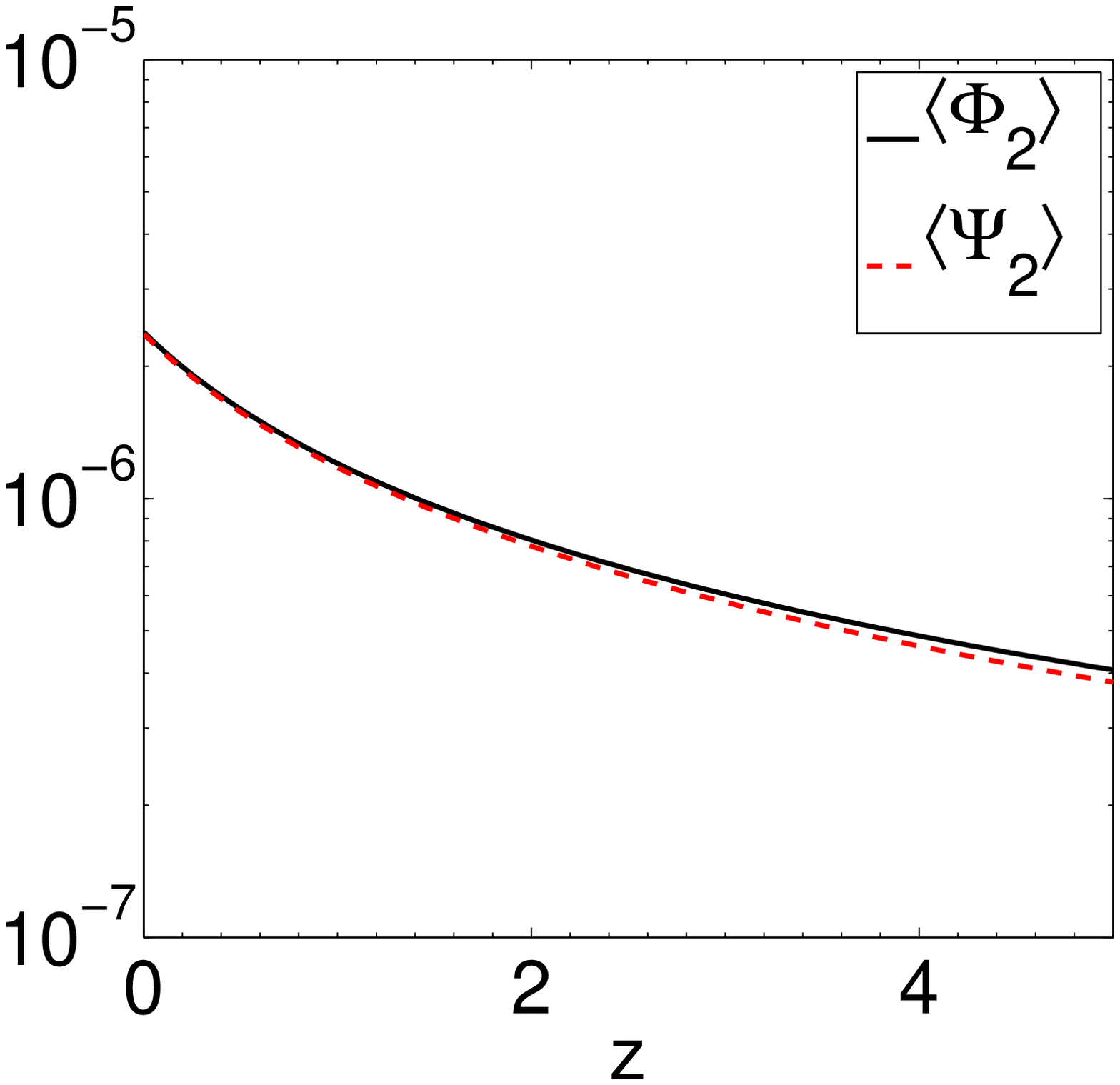}
\includegraphics[width=0.4\columnwidth]{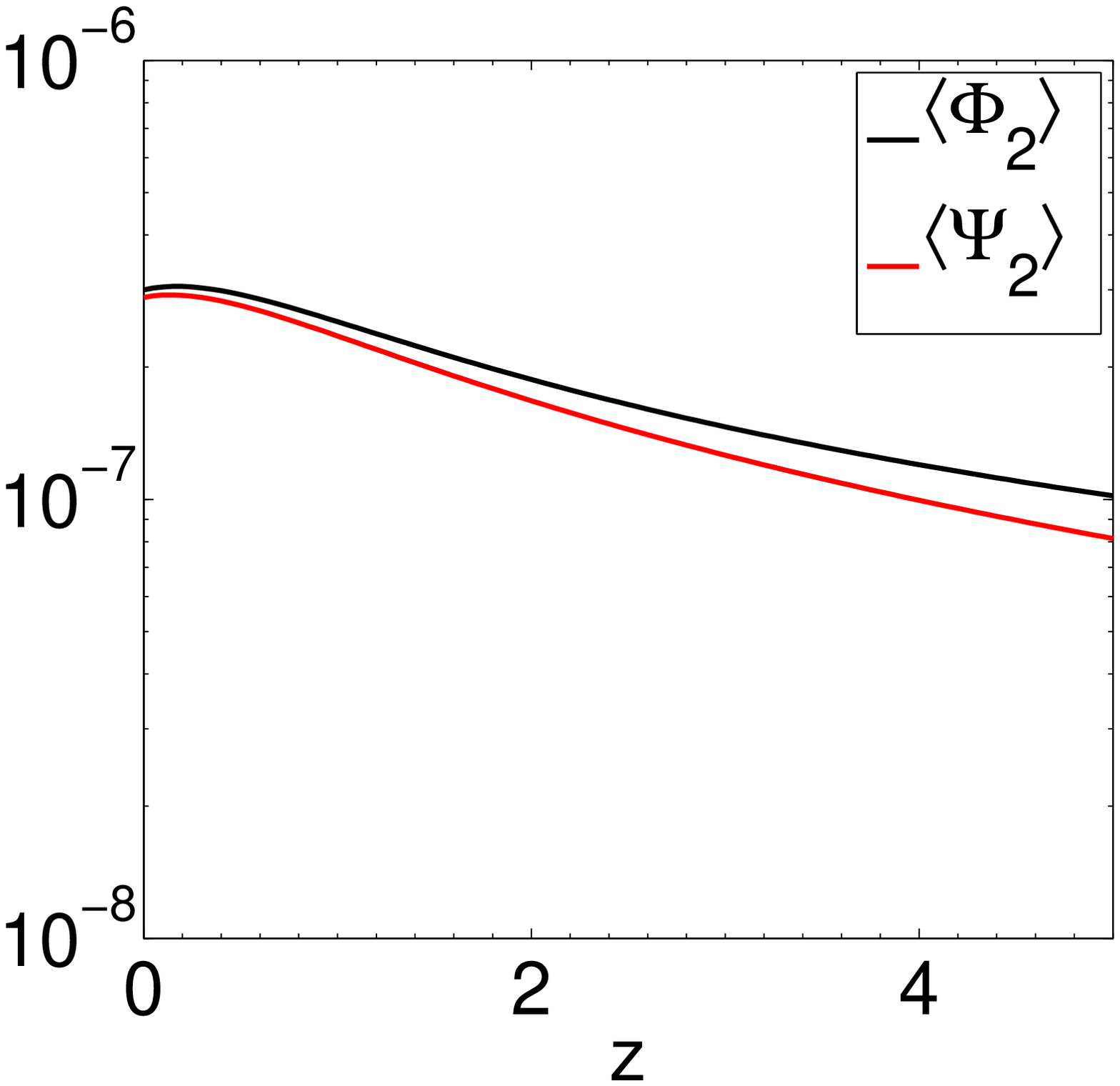}
\includegraphics[width=0.4\columnwidth]{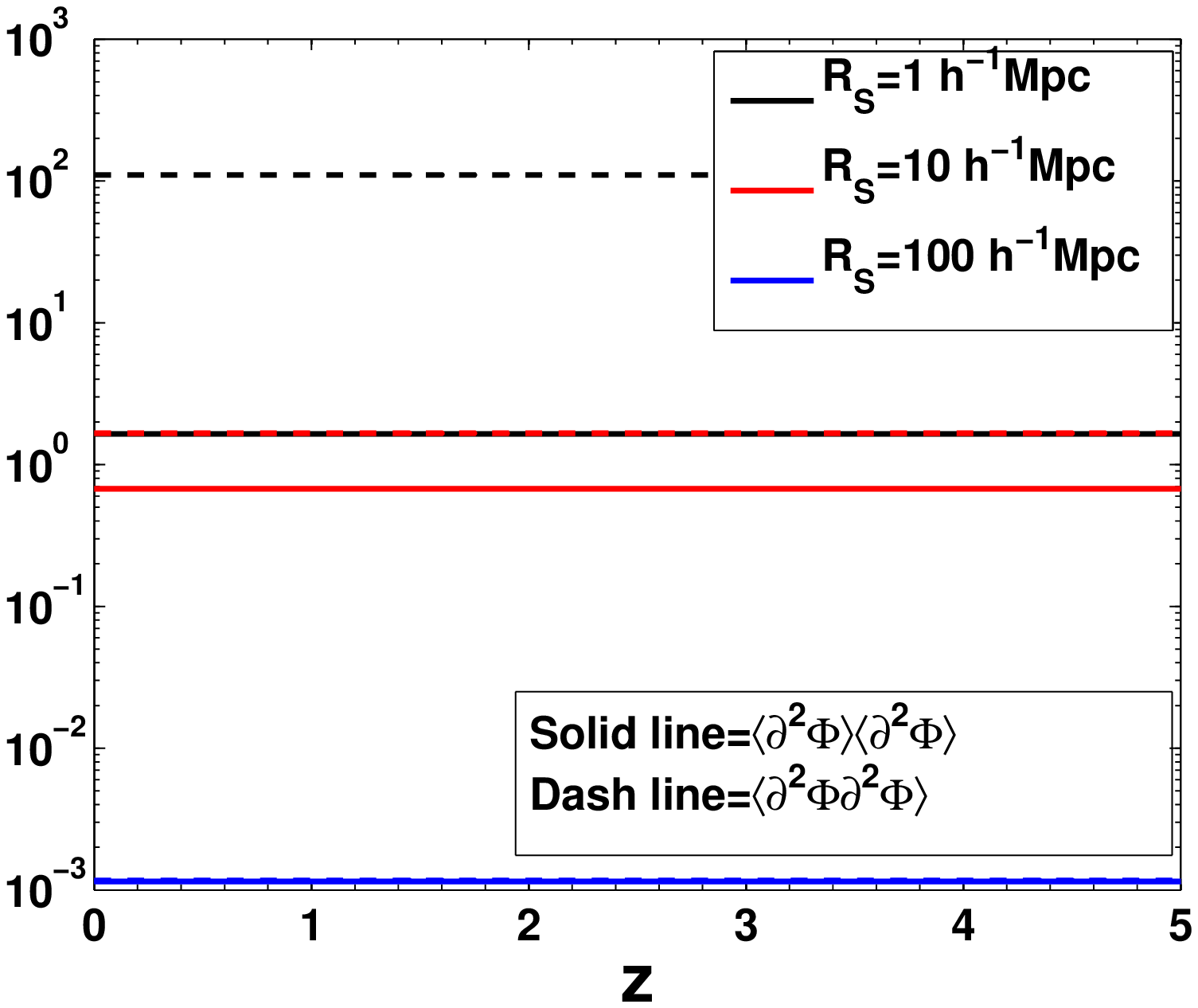}
\includegraphics[width=0.4\columnwidth]{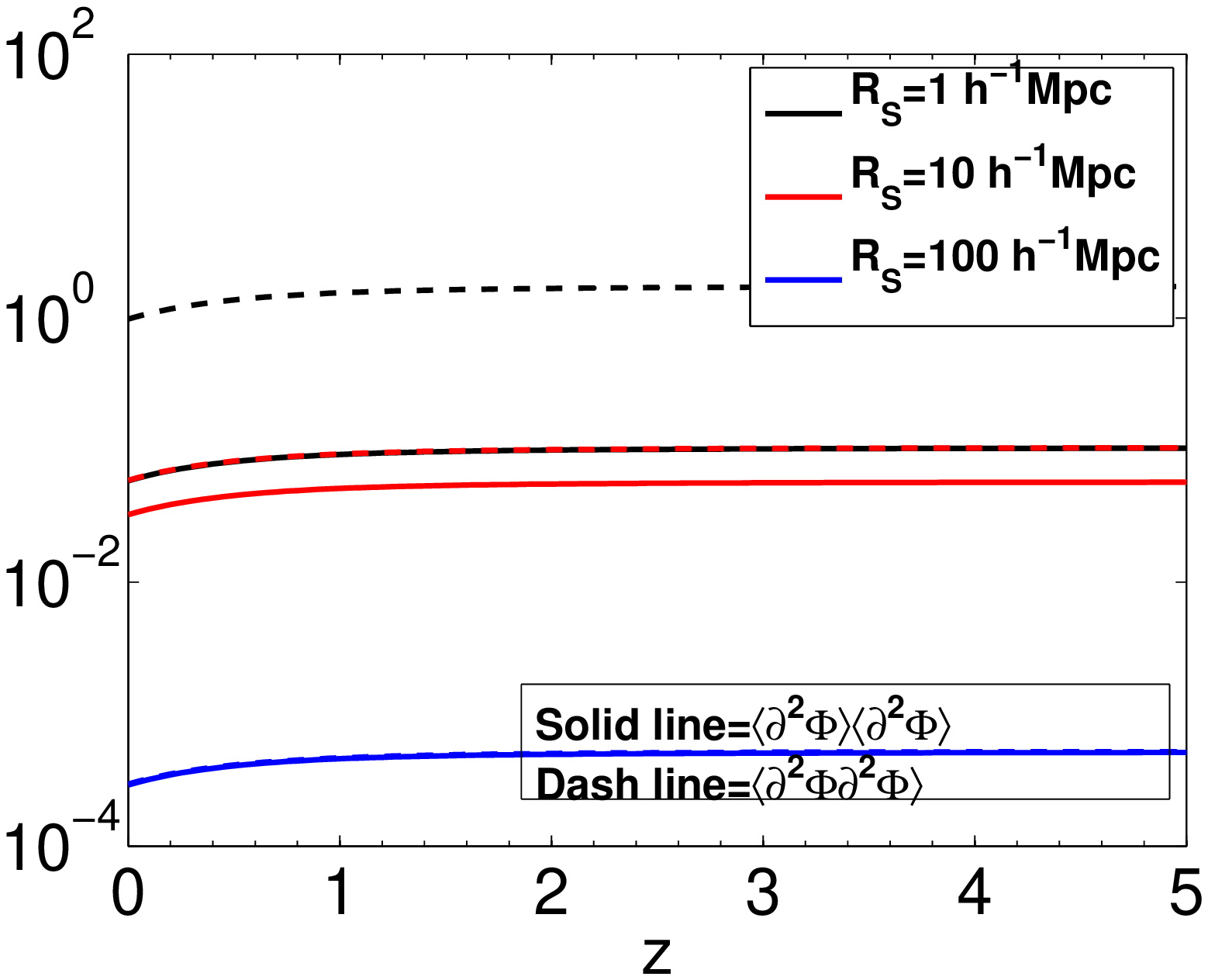}
\caption{The amplitude of the ensemble averages of various second-order terms which appear in the Hubble rate. }
\label{fig:phis}
\end{figure}

We show the ensemble averages of some of the second-order terms which appear in the  Hubble rates in Fig.~\ref{fig:phis}. Note that we also show the result of $\saverage{\partial^2 \Phi\partial^2 \Phi}$ for comparison.

\subsection{Comparison between the different definitions}

We can now turn to estimating and comparing the Hubble rates as well as their intrinsic variances as defined above consistently up to second order in perturbation theory. When determining the ensemble average of the Hubble rate, we shall consider two alternatives: a kinematical ensemble average given by $\overline{H}_\mathcal{D}$, and a dynamical one, which arises from taking the ensemble average of the Friedman equation: $\sqrt{\overline{H_\mathcal{D}^2}}$. We shall find that the difference between these two is large because the variance is large.

\begin{figure}[htb!]
\includegraphics[width=0.49\columnwidth]{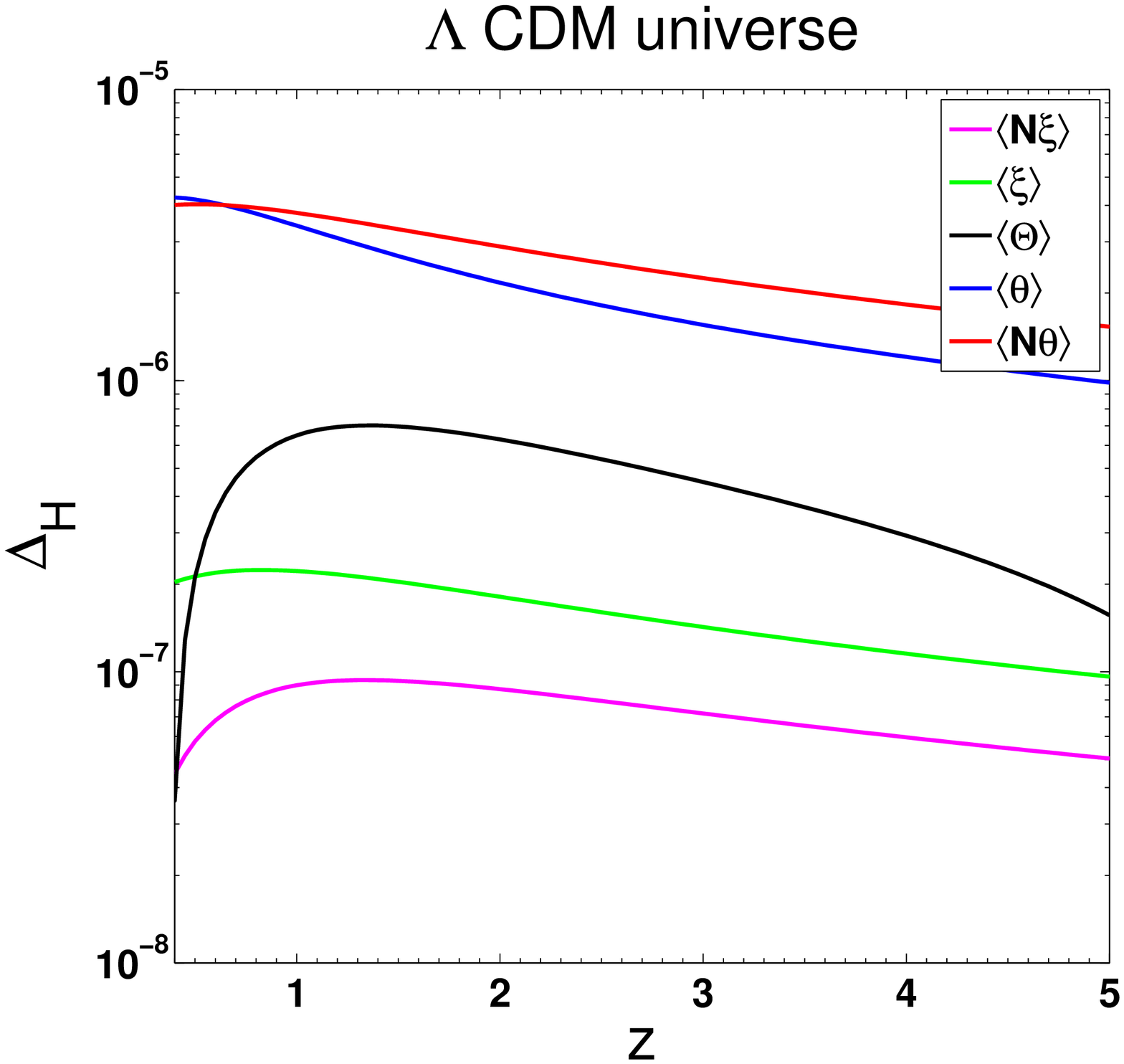}
\includegraphics[width=0.49\columnwidth]{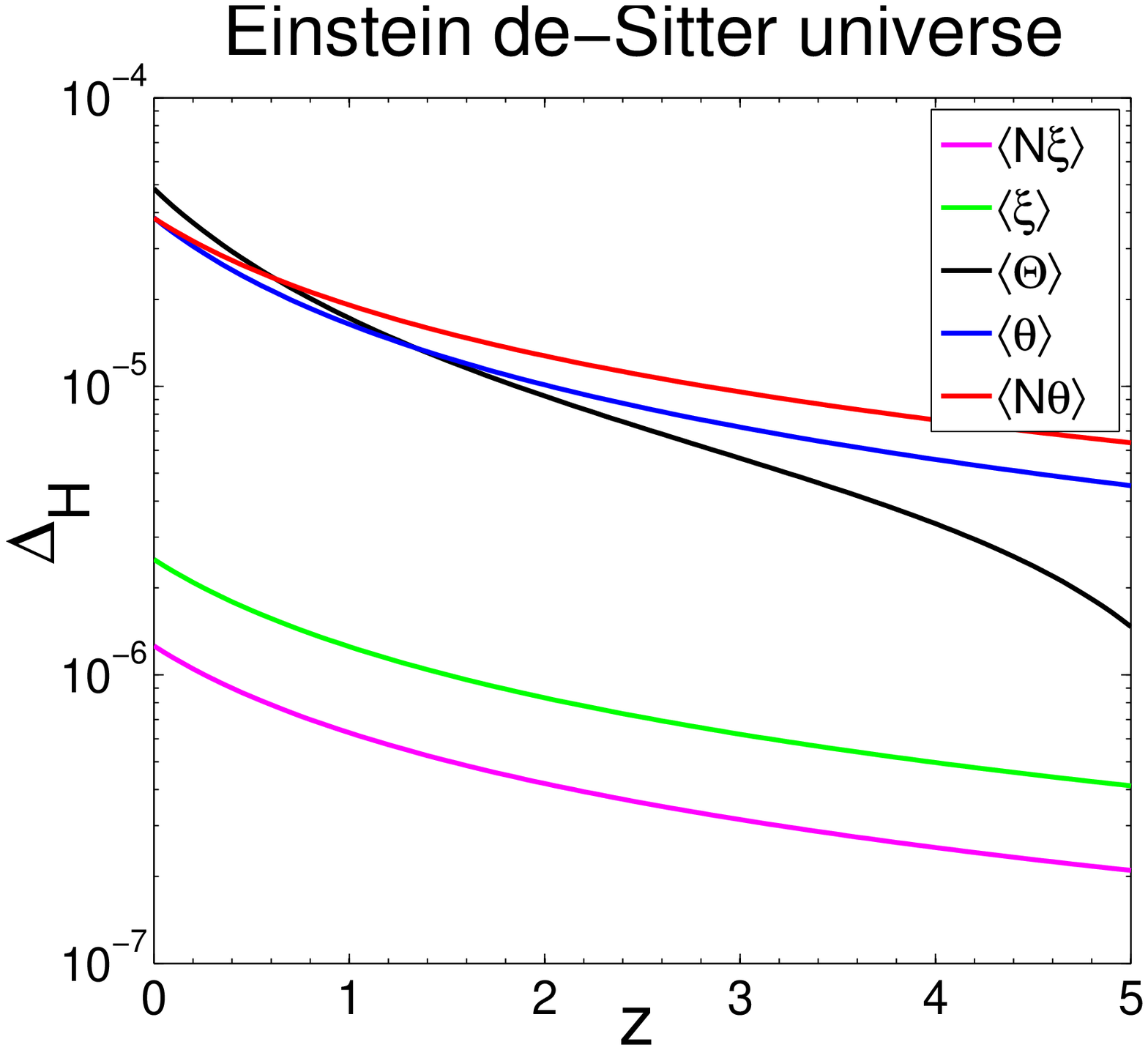}
\includegraphics[width=0.49\columnwidth]{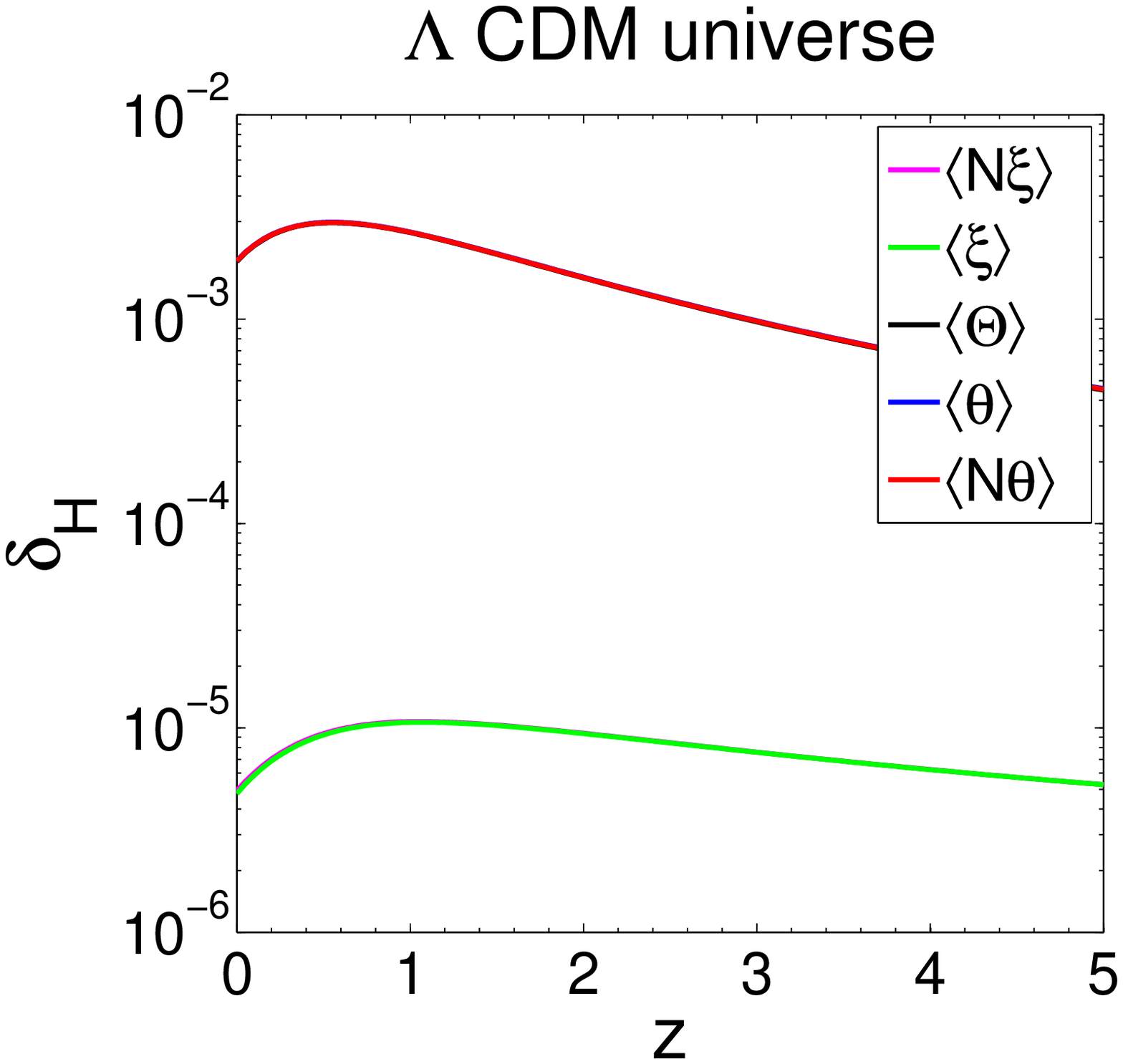}
\includegraphics[width=0.49\columnwidth]{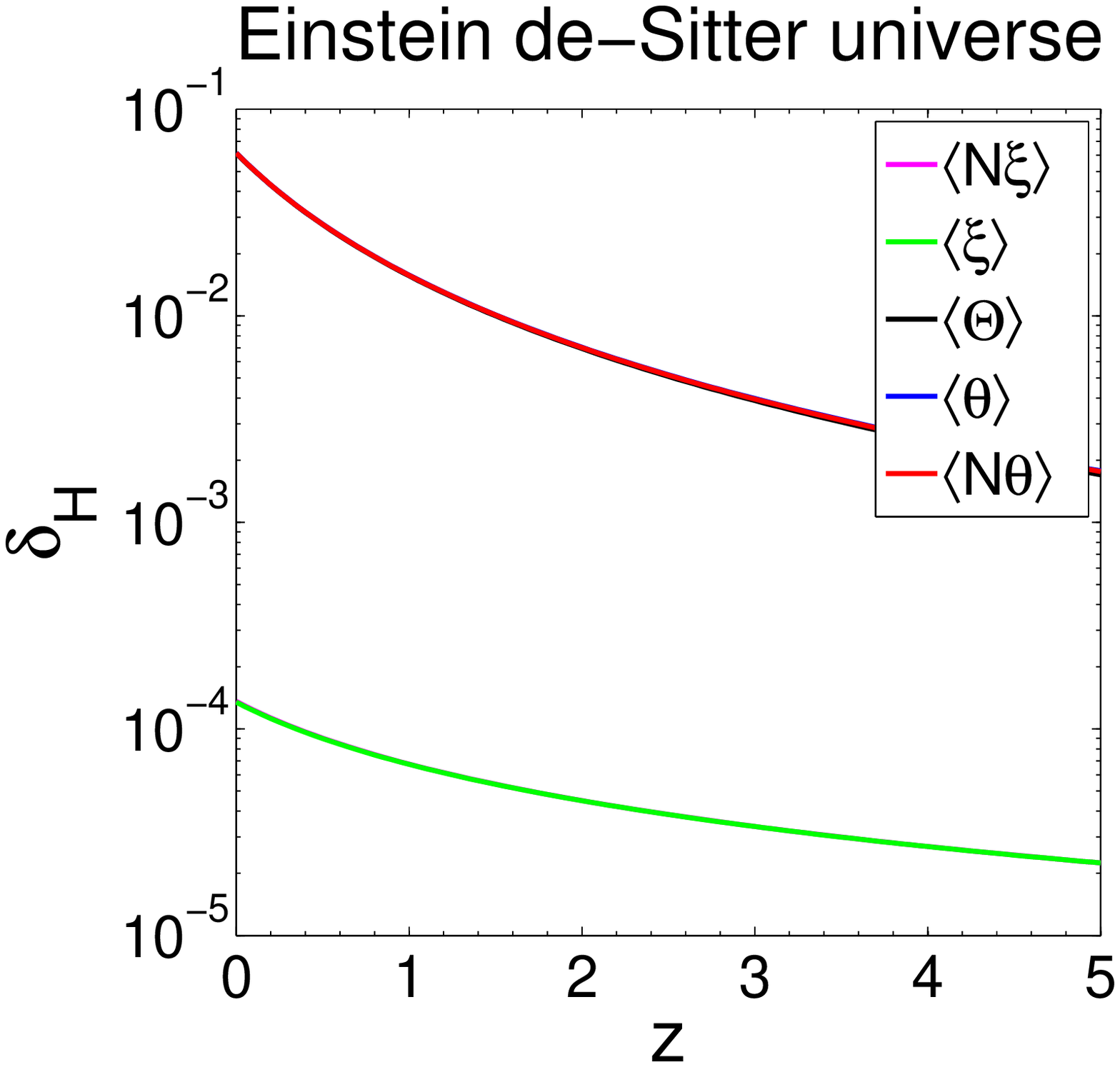}
\caption{Fractional change to the background Hubble rate as a function of redshift for the different definitions of averaged Hubble rates under study. Here we have averaged at the equality scale. Here, $\Delta_H=(\overline{H}_\mathcal{D}-H_0)/H_0$, and $\delta_H=(\sqrt{\overline{H_\mathcal{D}^2}}-H_0)/H_0$.}
\label{chap3Figure3}
\end{figure}

\begin{figure}[htb!]
\includegraphics[width=0.49\columnwidth]{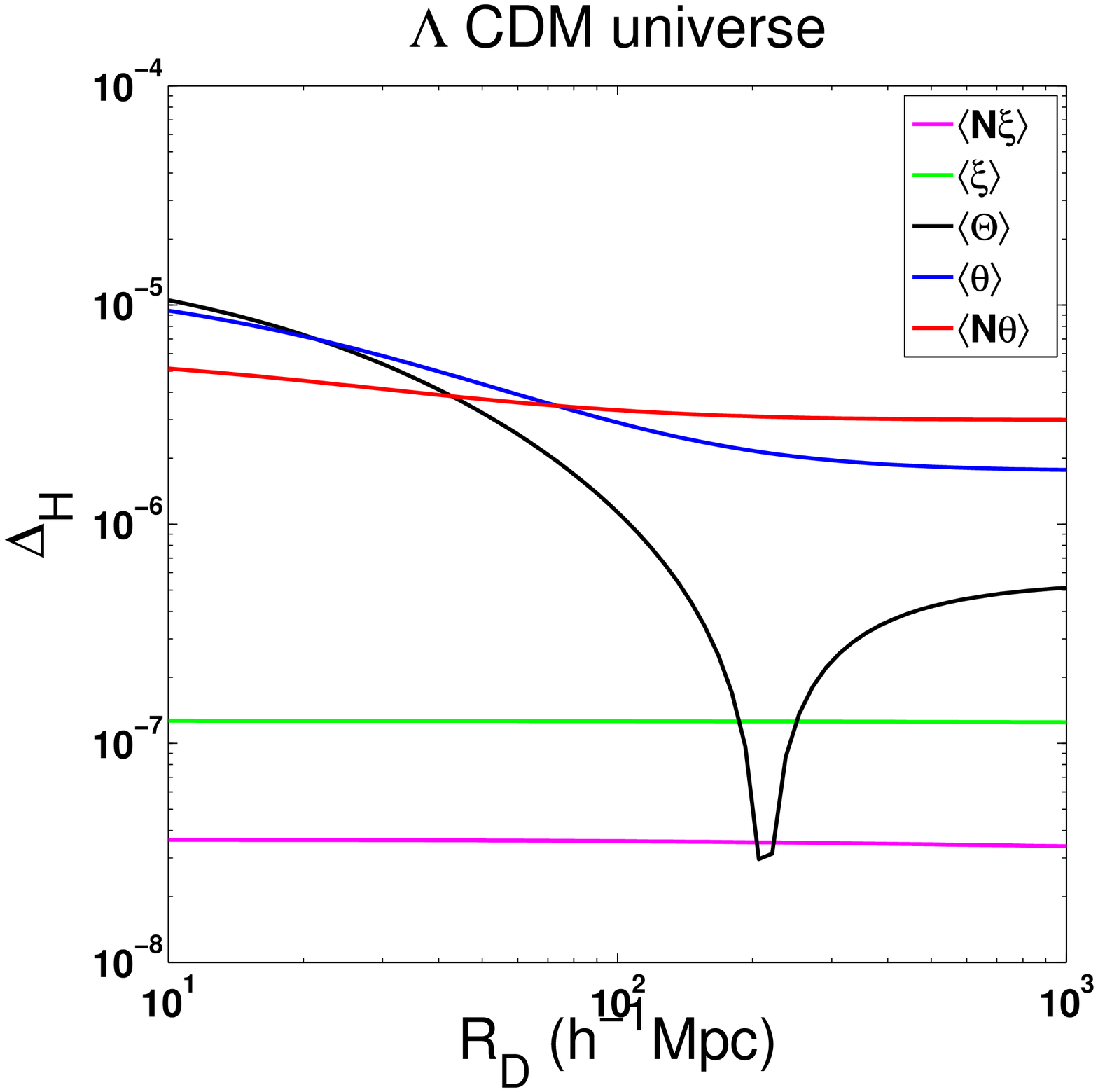}
\includegraphics[width=0.49\columnwidth]{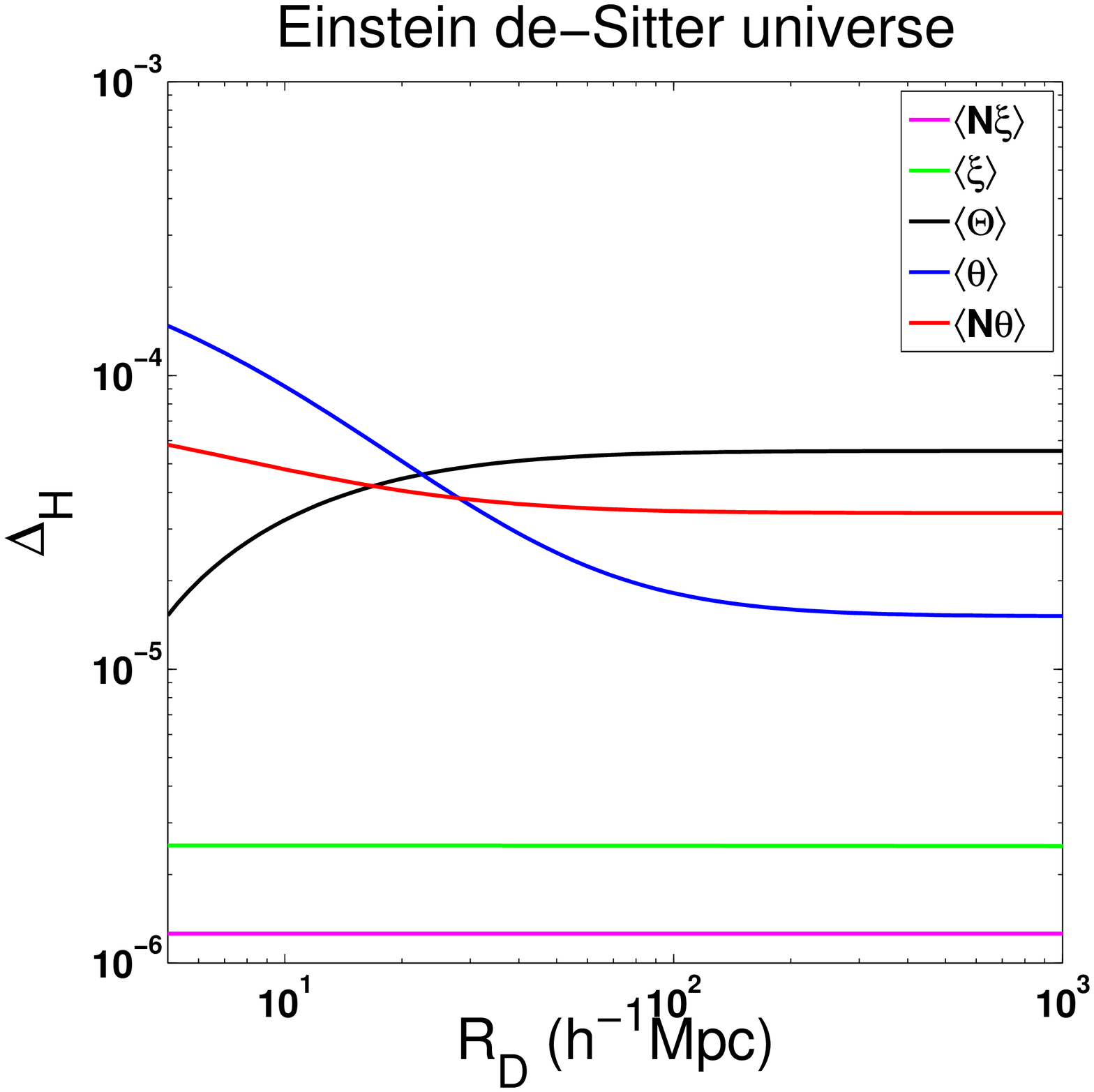}
\includegraphics[width=0.49\columnwidth]{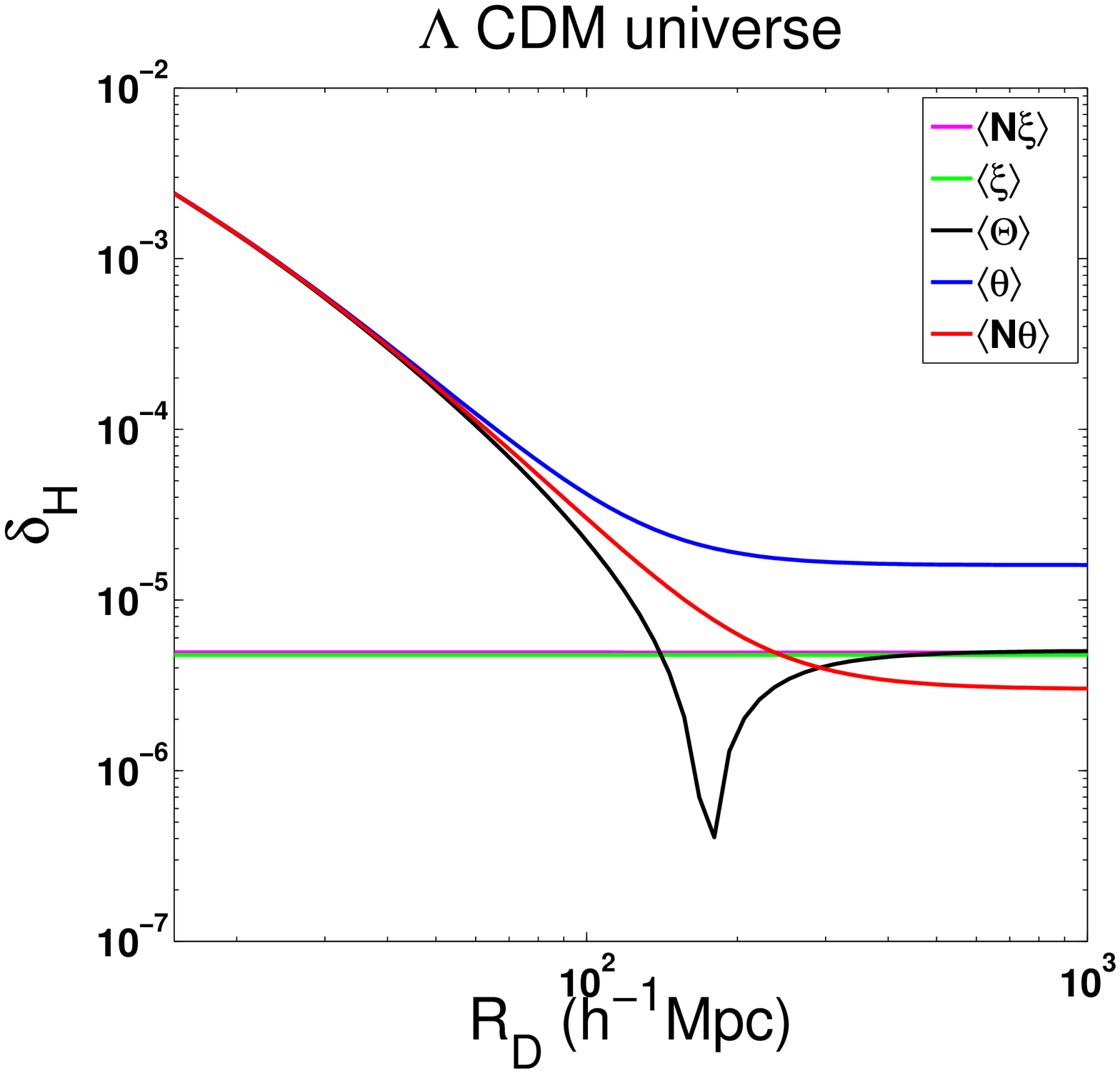}
\includegraphics[width=0.49\columnwidth]{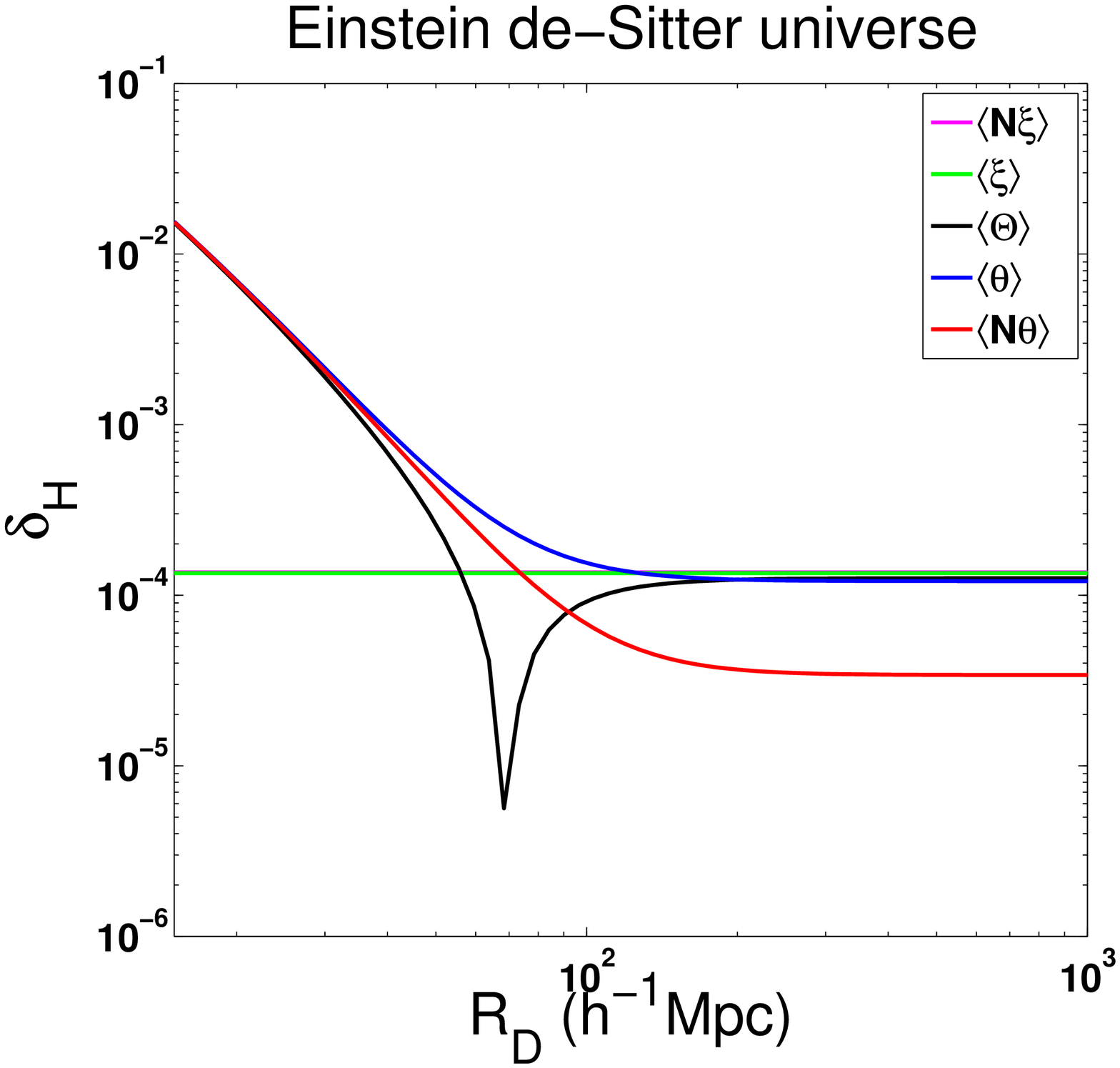}
\caption{Fractional change to the background Hubble rate as a function of the averaging scale for the different definitions of averaged Hubble rates under study. In both models, $\average{\xi}$ and $\average{N\xi}$ almost coincide and are indistinguishable in the figure. Notice the turn-down in $\langle{\Theta\rangle}_{\mathcal{F}}$  definition in Einstein de Sitter Universe, the physical meaning of this is not immediately obvious}
\label{chap3Figure4}
\end{figure}

\begin{figure}[htb!]
\includegraphics[scale=0.350]{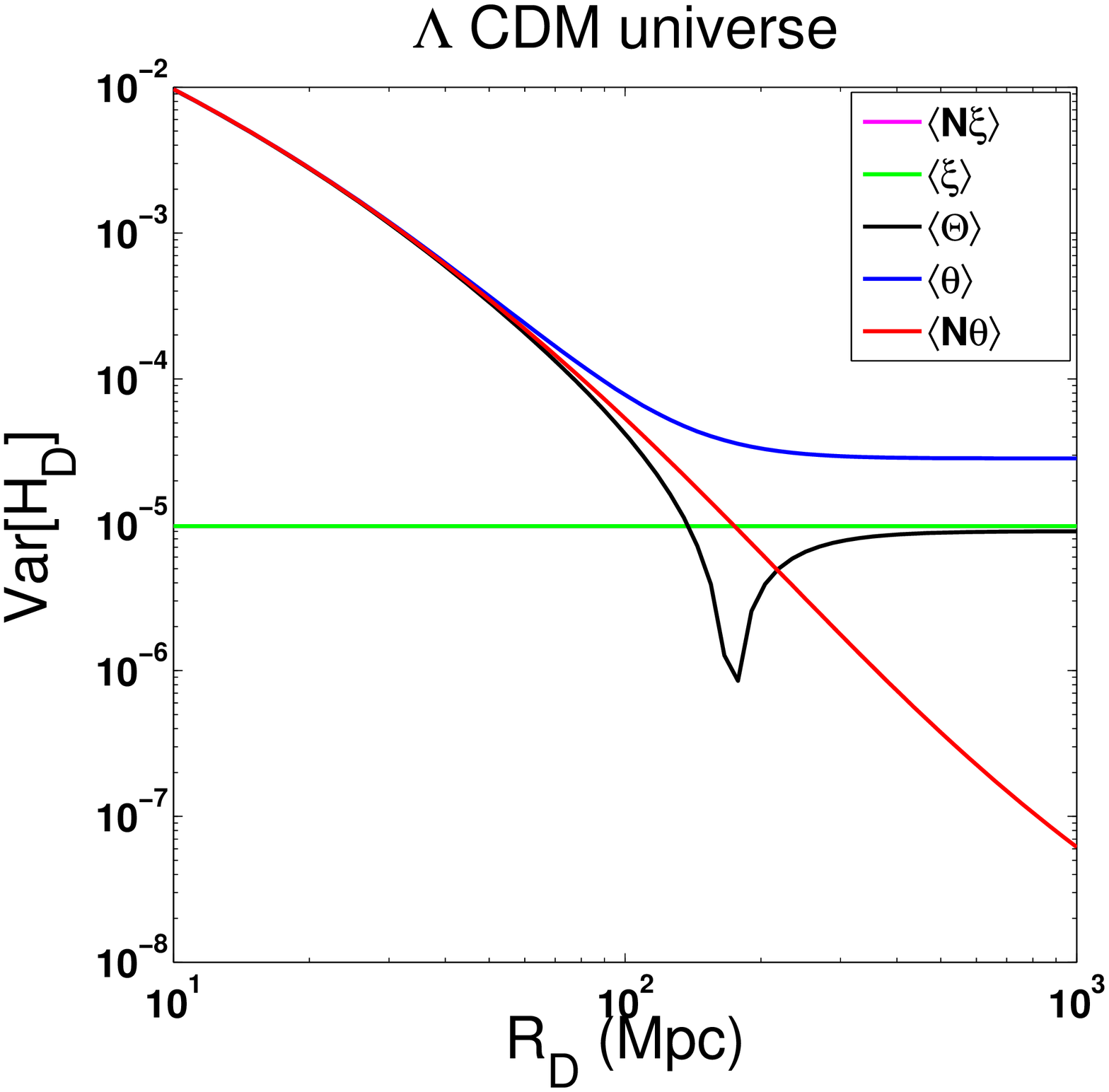}
\includegraphics[scale=0.350]{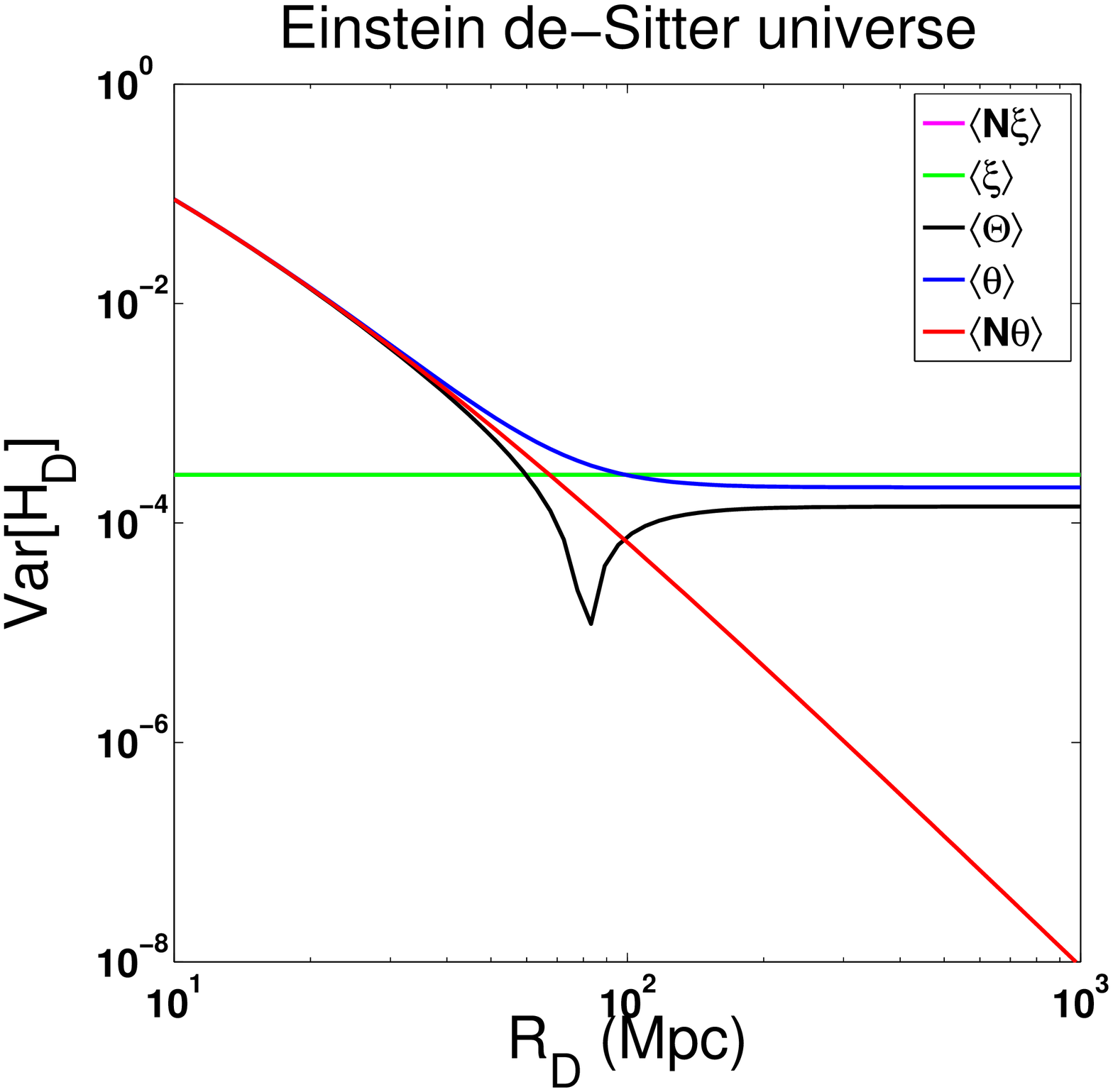}
\caption{Variance in the fractional change to the background Hubble rate as a function of redshift for the different definitions of averaged Hubble rates under study.}
\label{chap3Figure6} 
\end{figure}

Fig.~\ref{chap3Figure3} presents the evolution of the averaged Hubble rate as functions of redshift for different definitions of Hubble rate in a $\Lambda$CDM and an EdS scenarios. Fig.~\ref{chap3Figure4} depicts the values of the same Hubble rates at $z=0$ as functions of the averaging scale $R_{D}$, and Fig.~\ref{chap3Figure6} shows the scaling of their variances with the averaging scale $R_{D}$.

It is clear that the two types of Hubble rates defined here, i.e. those of the gravitational frame, and the ones defined in terms of the physical matter flow can be distinguished as far as the magnitude of their mean  and  variance are concerned.

First, the ones defined through the local expansion of the observers' worldlines, $\average{\xi}$ and $\average{N\xi}$ present a very small  correction to the FLRW background Hubble rate,  which is of the order $10^{-5}$ for $\Lambda$CDM and $10^{-4}$ for an EdS scenario  at 20 $ h^{-1}$Mpc .  Such a small effect was also reported from a study using  numerical simulation \cite{Zhao:2009yp}. Moreover, they appear to  be  scale independent when compared with the definition based on $\theta$.  
The scale dependence is determined by terms with two angle brackets $\saverage{\Phi}\saverage{\Phi}$.  For  the definitions based on  $\xi$,  for example,  the variance in $\average{\xi}$ or $\average{N\xi}$ is scale dependent but its scale dependence is suppressed when compared with the definitions based on $\theta$ because the dominant term in $\xi$ definition is $\saverage{\Phi}\saverage{\Phi}$ which is of the order of $10^{-10}$ at 20 $h^{-1}$Mpc  while in $\theta$ definition the dominant term is $\saverage{\partial^2 \Phi}\saverage{\partial^2\Phi}$ which is of the order of $10^{-4}$ at 20 $  h^{-1}$Mpc. 

Second, the Hubble rates defined through the local expansion of the matter worldlines systematically present corrections to the background Hubble rate, which is  two orders of magnitude bigger than the  previous ones, and are indistinguishable from each other, except when the averaging scale is much larger than the equality scale.   It is interesting to note that both the values of these averaged Hubble rates and their variances are indistinguishable up to scales of averaging of order $>$100 Mpc, after which they start  to differ. This scale have been interpreted in a previous work \cite{Clarkson:2009hr} as naturally defining the scale of statistical homogeneity of the universe (note that this is the case even for EdS; so it is not simply the equality scale). Around the same scale the expansion rate of the gravitational frame becomes comparable with the others because the peculiar velocity tends to zero. 

\begin{figure}[htb!]
\includegraphics[width=\columnwidth]{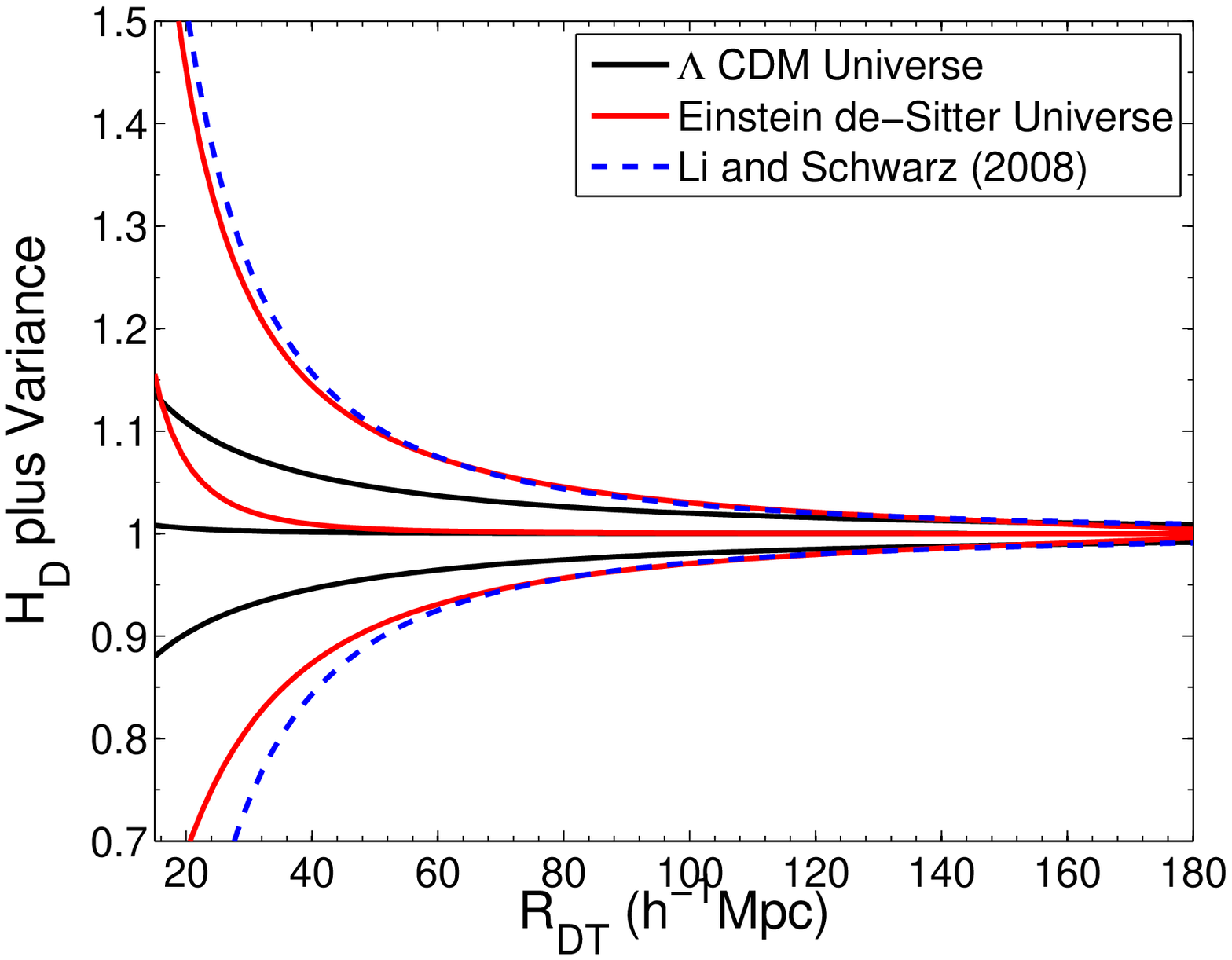}
\caption{Dynamical Hubble rate $\langle{\Theta\rangle}_{\mathcal{F}}$ today, plus/minus the variance as a function of the averaging scale (normalized to the background Hubble rate), where we have used a top-hat window function to define our domain for comparison with~\cite{Li:2007ny,Li:2008yj}. The blue curve represents the variance calculated in \cite{Li:2007ny,Li:2008yj}; differences for small domains are a consequence of the different transfer functions used here. }
\label{chap3Figure10}
\end{figure}

Finally, let us note that the results are consistent, for a pure CDM Universe, with those found on small scales in \cite{Li:2007ny,Li:2008yj}. This can be seen on Fig.~\ref{chap3Figure10}.

This analysis shows that the averaged Hubble rates defined through the expansion of the Newtonian-like or gravitational frame, as in \cite{Brown:2008ra,Brown:2009tg}, is not a good tracer of the expansion of the cosmic fluid, except beyond the homogeneity scale. The fluid frame is more relevant for local measurements since it is attached to the matter component of the Universe. The `gravitational frame', as we have referred to it here, seems useful on much larger scales, which is the situation in \cite{Brown:2008ra,Brown:2009tg} in which it was first evaluated~-- their domain was the Hubble scale. 

\subsection{Fluctuations in the measurement of $H_{0}$}

We would like to finish this paper by addressing the following questions:
\begin{itemize}
\item What is the physical relevance of the averaged Hubble rate and its variance?
\item Can there be any signature of backreaction in the observations leading to the measurement of $H_{0}$? 
\end{itemize}
First, let us note that on sufficiently small scales, such as scales smaller than $\sim100$ Mpc, which are the standard scales at which the Hubble rate is evaluated, and in a statistically homogeneous and isotropic Universe, spatial averages are expected to be a good approximation of what happens along the past lightcone on which observations are made. Along the past lightcone the monopole contribution to the Hubble rate, which is the one that remains once a full sky average has been performed, is exactly the covariant quantity $\Theta=\nabla_{a}u^{a}$ \cite{1966ApJ...143..379K}.

Hence, our estimate of $H_{\CD}$ on a scale $R_{\CD}$ can be interpreted as the average Hubble rate in a patch of the local Universe of size $R_{\CD}$ as long as this size remains sufficiently small compared with the Hubble scale. Moreover the variance we calculated is the intrinsic dispersion on the measurement of $H_{0}$ that comes from the fluctuations in the peculiar velocity of the sources and gravitational potential. In a concordance cosmology, this dispersion appears small, of order 1\% at a scale of $100$ Mpc, and even less on larger scales, as can be seen on Fig.~\ref{chap3Figure10}. 

This is consistent with previous estimates that were based on an estimate of the first order velocity power spectrum \cite{Wang:1997tp,1998ApJ...493..519S}. It is due to the fact that the pure second order terms cancel out consistently at second order in our expression of the variance, allowing only contributions of squares of first order quantities. As noted before, this is a similar effect to that found in \cite{Li:2007ny,Li:2008yj}, where the calculations were made in the comoving synchronous gauge, for a pure CDM Universe. Our calculation of $\langle\Theta\rangle_{\cal F}$ using the gauge-invariant approach of~\cite{Gasperini:2009mu}  corresponds to a gauge-invariant version of the average expansion rate in the synchronous gauge.

To quantify the backreaction effect on the variance for a large class of cosmological models, we provide a fitting formula for the variance of the Hubble rate (defined via the flow of matter), $Var[H]$, that is accurate to a few percents across the scales of interest:
\begin{eqnarray}
\log_{10}(Var[H]) &=& - 43.61 + 46.0 \Omega_m^{0.0293}- 0.7969 f_b^{0.0347}\\
 & &+ \lambda (\log_{10}R)^\alpha + \gamma \exp(-\beta(\log_{10}R)^2)\nonumber
\end{eqnarray}
where
\begin{eqnarray}
\lambda&=& 10.32-9.084\Omega_m^{0.0469}- 3.611/f_b^{0.00497}\\ \nonumber
\gamma&=& 1.309 - 2.355 \Omega_m^{0.055}- 1.073 f_b^{2.1778} \\ \nonumber
\beta&=&-1.805+ 3.260 \Omega_m^{0.0279}- 0.7180 f_b^{0.665}\\ \nonumber
\alpha&=& 1.222 + 0.0334 \Omega_m^{3.635} + 0.0591 f_b^{0.3944}.
\end{eqnarray}
This formula gives the variance on the measurement of $H_{0}$, normalised to the value of $H_{0}$ for $H_{\mathcal{D}}=\frac{1}{ 3}\average{N\theta}$ definition (The fitting formula for other definitions exist but they are more complicated and less accurate): $\Omega_{m}$ is the CDM density parameter, $f_{b}$ the baryon fraction, and $R$ the length characteristic of the survey, i.e. the distance to the farthest object (in units of Mpc). Note that this fitting formula is valid for a Gaussian window function. Top-hat window functions generically lead to a slight increase of the variance.

\section{Conclusion}\label{sec4}

In this work, we have presented the first comparison of  the different definitions of the averaged Hubble rate that can be found in the literature.  We did this by calculating these various  definitions consistently to second order  in cosmological  perturbation theory. Also for the first time we have calculated the average of the expansion rate using the formalism of~\cite{Gasperini:2009mu}, which the authors claim  is gauge invariant at some limit. We have also  found that the  definitions that involve the flow of the dust matter component are consistent with each other at second order in cosmological perturbation theory, but differ significantly on small scales from the  definition based one the expansion of the  coordinate  grids. In particular, we noticed the following features of the averaged Hubble rate:


\begin{itemize}
\item On small scales all definitions which involve the matter flow agree, and give a small sub-percent change to the background Hubble rate.
\item The  variance in the average of the expansion rate of the gravitational frame  are  very small and appear to exhibit weak  scale dependence when compared with the definition involving the matter flow. 
\item On large scales (much larger than the equality scale) all definitions become scale invariant once their ensemble average is evaluated.
\item The hypersurface used in averaging is not really important when computing the variance for perturbed FLRW model, the differences only  show up on large scales and it is only noticeable  in Einstein de Sitter models.
\item Including $N$ in the definition of the averaged expansion  leaves a residual effect on large scales, and it tends to reduce the backreaction effect.   The inclusion of the lapse function is made compulsory if one wants to keep the coordinate time $t $ as the proper time in the averaged model (there is a discussion on this issue in \cite{Clarkson:2009hr}). But this is only one possible choice, since no-one knows how to explicitly construct the average model in this setting where only the scalars are averaged. 
\end{itemize}

We have also derived the dispersion affecting the Hubble rate and arising from the peculiar velocities of the matter flow. We found an effect consistent with previous estimates from backreaction in the literature \cite{Li:2007ny,Li:2008yj}, and our results are consistent with effects evaluated previously \cite{Wang:1997tp, Clarkson:2009hr}. 


We close with a comment on the origin of the scale dependence of the various quantities. The scale dependence we have found here  comes only from `non-connected'  terms such as $\saverage{\Phi}\saverage{\partial^2\Phi}$ since the domain size factors out of all other terms (for details on how this type of terms appear see \cite{Clarkson:2009hr}). 
Non-connected terms only arise when we perform averages in the spacetime itself, which many authors on backreaction have stressed is important. It is interesting to note that these terms (i.e those ones involving  laplacian of a gravitational potential, for example $\saverage{\Phi}\saverage{\partial^2\Phi}$) do not appear if we treat perturbations as fields propagating on the background, and calculate average quantities only with respect to the background geometry i.e., if we perform a Euclidean average and not a Riemannian one~\cite{Baumann:2010tm}.

\section*{Acknowledgments}
We thank  George Ellis, Giovanni Marozzi, Cyril Pitrou, Syksy R\"as\"anen, Dominik Schwarz  and  Alex Wiegand for helpful discussions.  We also thank the anonymous referee for suggestions that improved the quality of this paper. JL is supported by the Claude Leon Foundation (South Africa). OU is supported by National Institute for Theoretical Physics (NITheP), South Africa and the University of Cape Town. CC is funded by the NRF (South Africa)

\appendix

\begin{widetext}

 \section{Second-order perturbation theory}\label{appendA}

 The Poisson gauge is particularly elegant for scalar perturbations because with $n^a$ defined orthogonal to the spatial metric $h_{ij}$, the electric and the magnetic part of the Weyl tensor becomes
\begin{eqnarray}
E^{(n)}_{ij}&=&\frac{1}{2}\left(h_i^{~a}h_{j}^{~b}-\frac{1}{3}h_{ij}h^{ab}\right)\left\{\tilde\nabla_{a}\tilde\nabla_{b}\left[\Phi+\Psi-\Phi^2-\Psi^2+\frac{1}{2}\left(\Phi^{(2)}+\Psi^{(2)}\right)  \right]+\tilde\nabla_a\Phi\tilde\nabla_b\Phi-\tilde\nabla_a\Psi\tilde\nabla_b\Psi  \right\}\\
H^{(n)}_{ij}&=&0.
\end{eqnarray}
In the rest frame $n^a$, then, the gravitational field is silent, and, with $\Psi=\Phi$ is a pure potential field. Hence, $n^a$ may be considered as the rest-frame of the gravitational field, or the Newtonian-like frame, and so defines natural hypersurfaces with which to perform our averages. By contrast, in the frame $u^a$ the Weyl tensor has non-zero $H_{ab}$~\cite{Clarkson:2009hr}.

The Einstein Equation for a single fluid with zero pressure and no anisotropic stress $\Psi=\Phi$, and $\Phi$ obeys the Bardeen equation
\begin{equation}
\Phi''+ 3 \mathcal{H} \Phi' +  a^2\Lambda  \Phi= 0 =\ddot\Phi+4H\dot\Phi+\Lambda\Phi\,. 
\label{equation of motion for
Bardeen potential}
\end{equation}
and $'=d/d\eta$, and $\mathcal{H}=a'/a$ is the conformal Hubble rate. 
All first-order quantities can be derived from $\Phi$.
 The solution to the growing mode of the Bardeen equation may be written as
\begin{equation}
\Phi(\eta,\bm x)=g(\eta)\Phi_0(\bm x)
\end{equation}
where $\Phi_0(\bm x)$ is the Bardeen potential today ($\eta=\eta_0,~z=0$) and $g(\eta)$ is the growth suppression factor, which may be approximated, in terms of redshift, as~\cite{Lahav:1991wc,Carroll:1991mt}
\begin{eqnarray}\label{gfac}
g(z)& =& \frac{5}{2} g_{\infty} \Omega_{m}(z) \left\{
\Omega_{m}(z)^{4/7} - \Omega_\Lambda(z)+ \left[ 1 + \frac{1}{2}
\Omega_{m}(z)\right] \left[1 + \frac{1}{70} \Omega_\Lambda(z)
\right] \right\}^{-1}.
\end{eqnarray}
and $g_\infty$ is chosen so that $g(z=0)=1$. 
 
The second-order solutions for $\Psi^{(2)}$ and $\Phi^{(2)}$ are given by~\cite{Bartolo:2005kv}. We quote their results directly:
\begin{eqnarray}
\label{PSI}
\Psi^{(2)}(\eta,\bm x)&=&\left( 
B_1(\eta)-2g(\eta)g_{m} -\frac{10}{3}(a_{\rm nl}-1)g(\eta)g_{m}
\right)\Phi_0^2 
+\left( B_2(\eta) -\frac{4}{3}g(\eta)g_{m}  \right) \Bigg[ \nabla^{-2} 
\left( \partial^i \Phi_0 
\partial_i \Phi_0 \right) \Bigg.
\nonumber\\ & &  
\Bigg.- 3 \nabla^{-4} \partial_i \partial^j
\left(\partial^i \Phi_0 \partial_j \Phi_0 \right) \Bigg]+ B_3(\eta) \nabla^{-2} \partial_i\partial^j(\partial^i \Phi_0 \partial_j \Phi_0 )+B_4(\eta) \partial^i \Phi_0 \partial _i\Phi_0 \, ,\\
\label{PHI}
\Phi^{(2)}(\eta,\bm x)&=&\left( B_1(\eta)+4g^2(\eta)
-2g(\eta)g_{m} -\frac{10}{3}(a_{\rm nl}-1)g(\eta)g_{m}
\right)\Phi_0^2 
+\Bigg[ B_2(\eta)+\frac{4}{3} g^2(\eta) \left( e(\eta)+\frac{3}{2} \right)
-\frac{4}{3}g(\eta)g_{m} \Bigg]  \nonumber \\
&&
 \times \Bigg[ \nabla^{-2} 
\left( \partial^i \Phi_0 
\partial_i \Phi_0 \right) - 3 \nabla^{-4} \partial_i \partial^j
\left(\partial^i \Phi_0 \partial_j \Phi_0 \right) \Bigg] 
+B_3(\eta) \nabla^{-2} \partial_i\partial^j(\partial^i \Phi_0 \partial_j 
\Phi_0 )+B_4(\eta) \partial^i \Phi_0 \partial _i\Phi_0\, ,
\end{eqnarray}
where 
$B_i(\eta)={\mathcal H}_0^{-2} \left(f_0+3 \Omega_{0}/2 \right)^{-1} 
\tilde{B}_i(\eta)$ with the following definitions
\begin{eqnarray}
\label{B1B2}
\tilde{B}_1(\eta)=\int_{\eta_m}^\eta \d\tilde{\eta} \,{\mathcal H}^2(\tilde{\eta}) 
(f(\tilde{\eta})-1)^2 C(\eta,\tilde{\eta})\mbox{,}\,\,\,\,\,\,\,\,\,
\tilde{B}_2(\eta)=2\int_{\eta_m}^\eta \d\tilde{\eta} \, {\mathcal H}^2(\tilde{\eta}) 
\Big[2 (f(\tilde{\eta})-1)^2-3
+3 \Omega_m(\tilde{\eta}) \Big] C(\eta,\tilde{\eta})\, , \\
\tilde{B}_3(\eta)=\frac{4}{3} \int_{\eta_m}^\eta \d\tilde{\eta} \left(e(\tilde{\eta})
+\frac{3}{2} \right) C(\eta,\tilde{\eta})\mbox{,} \,\,\,\,\,\,\,\,\,\,\,\,\,\,\,\,\,\,\,\,\,\,\,\,\,\,\,\,\,\,\,\,\,\,\,\,\,\,\,
\tilde{B}_4(\eta)= - \int_{\eta_m}^\eta \d\tilde{\eta} \,C(\eta,\tilde{\eta},)\,\,\,\,\,\,\,\,\,\,\,\,\,\,\,\,\,\,\,\,\, \,\,\,\,\,\,\,\,\,\,\,\,\,\,\,\,\,\,\,\,\,\,\,\,\,\,\,\,\,\,\,\,\,\,\,\,\,\,\,\,\,\,\,\,\,\,\,\,
\end{eqnarray}
and 
\begin{equation}
C(\eta,\tilde{\eta})= g^2(\tilde{\eta}) a(\tilde{\eta}) 
\Big[ g(\eta){\mathcal H}(\tilde{\eta})-g(\tilde{\eta}) 
\frac{a^2(\tilde{\eta})}{a^2(\eta)} {\mathcal H}(\eta) \Big] \, ,
\end{equation}
with $e(\eta)=f^2(\eta)/\Omega_m(\eta)$ and
\begin{equation}
f(\eta)=1+\frac{g'(\eta)}{{\mathcal H} g(\eta)}\, .
\end{equation}
$g_m$ denotes the value of $g(\eta_m)$, deep in the matter era before the cosmological constant was important. We also have $a_{\rm nl}$ which denotes any primordial non-Gaussianity present. We set this to unity, representing a single field slow-roll inflationary model. For details on how the  spatial average  of the second order Bardeen Potential may be evaluated see~\cite{Clarkson:2009hr}

\section{Hubble rates}

In this appendix, we present the different Hubble rates, consistently at second order. The superscript determines the quantity that has been averaged to define the average Hubble rate. 

\begin{eqnarray}
 H_\mathcal{D}^{N\xi} &=& H - \saverage{\dot \Phi}- 3 \saverage{\dot \Phi}\saverage{\Phi}+ \saverage{\Phi \dot \Phi}-\frac{1}{2}\saverage{\Psi_2}\mbox{ .}
 \end{eqnarray}

 \begin{eqnarray}
 H_\mathcal{D}^{\xi}&=& H -\saverage{\dot \Phi}- H \saverage{\Phi}-3\saverage{\dot \Phi}\saverage{\Phi}+2\saverage{\Phi\dot \Phi}-3H\saverage{\Phi}^2+ \frac{9}{2}\saverage{\Phi^2}-\frac{1}{2}\saverage{\dot \Psi_2}-\frac{1}{2}H \saverage{\Phi_2}\mbox{ .}
\end{eqnarray}

 \begin{eqnarray}
H_\mathcal{D}^{N\theta}&=& H - \saverage{\dot \Phi}-3 \saverage{\dot \Phi}\saverage{\Phi} + 2 \saverage{\dot \Phi\Phi}-\frac{1}{2}\saverage{\dot \Psi_2}+ \frac{\left( 1+z\right)}{6}  \partial_kv^k_2 - \frac{2\left( 1+z\right) ^2}{9\Omega_{m} H^2}\left[ \saverage{ \partial^2\dot \Phi}+ H \saverage{ \partial^2\Phi}\right]\nonumber\\
 & &+ \frac{\left( 1+z\right) ^2}{\Omega_{m}^2 H^3}\left[ \frac{8}{9}H \left( 1+\frac{\Omega_{m}}{2}\right) \saverage{ \partial_k\dot \Phi  \partial^k\Phi}+ \frac{4}{9}H^2\left( 1+\Omega_{m}\right) \saverage{ \partial_k \Phi  \partial^k \Phi} - \frac{4}{9}\saverage{ \partial_k \dot \Phi \partial^k\dot \Phi}\right]\nonumber\\
 & &+ \frac{2\left( 1+z\right) ^2}{3\Omega_{m} H^2}\left[ \frac{2}{3}\saverage{\Phi \partial^2\dot \Phi}+ \frac{2}{3}H\saverage{\Phi \partial^2 \Phi}- \saverage{ \partial^2\dot \Phi}\saverage{\Phi}- H \saverage{ \partial^2 \Phi}\saverage{\Phi}\right]\mbox{ .}
 \end{eqnarray}

\begin{eqnarray}
  H_\mathcal{D}^{\theta}&=& H - \saverage{\dot \Phi}-H \saverage{\Phi}- 3\saverage{\dot \Phi}\saverage{\Phi}+ 2\saverage{\Phi \dot \Phi}-3 H \saverage{\Phi}^2+ \frac{9}{2} H \saverage{\Phi^2} -\frac{1}{2}\left[\saverage{\dot \Psi_2}+2H \saverage{\Phi_2}\right] \nonumber\\
 & &
  +\frac{\left( 1+z\right)}{6}  \saverage{ \partial_kv^k} 
    -\frac{2\left( 1+z\right)^2}{9\Omega_{m} H^2}\left[ \saverage{ \partial^2\dot \Phi}
+ H \saverage{ \partial^2 \Phi}\right] 
+\frac{2\left( 1+z\right)^2}{9 \Omega_{m}^2 H^3}\left[ \saverage{ \partial_k\dot \Phi \partial^k\dot \Phi}
+2H \saverage{ \partial_k\dot \Phi  \partial^k\Phi}\right.\nonumber\\
 & &
\times\left. \left(1+\frac{3}{2}\Omega_m\right)+H^2\saverage{ \partial_k\Phi \partial^k\Phi}\left(1+3\Omega_m\right)\right]+\frac{2\left( 1+z\right)^2}{3 \Omega_{m} H^2}\left[\saverage{\Phi \partial^2\dot \Phi}\ - \saverage{ \partial^2\dot \Phi}\saverage{\Phi}\right.\nonumber\\
 & &
+\left. H \saverage{\Phi \partial^2\Phi} -H \saverage{ \partial^2\Phi}\saverage{\Phi}\right]\mbox{ .}
\end{eqnarray}
\begin{eqnarray}
 H_\mathcal{F}^{\Theta}&=&  H - \saverage{\dot \Phi}-\saverage{\Phi}H\left[ 1+\frac{3}{2}\Omega_m H g_I\right]-\frac{2(1+z)^2}{9\Omega_mH^2}\left[\saverage{\partial^2\dot\Phi}+H \saverage{\partial^2\Phi}\right]-2\saverage{\Phi \dot \Phi}+\frac{(1+z)}{6}\saverage{\partial_k v^k_2}\nonumber\\
 & & 
-\frac{1}{2}\left[ \saverage{\dot \Psi_2}- H\saverage{\Phi_2} \right] -3H\saverage{\Phi}^2\left[(1+g_IH)+ \frac{3}{2}\Omega_m Hg_I(1-Hg_I)\right]-3\saverage{\dot\Phi}\saverage{\Phi}\left[1+H g_I\right] \nonumber\\
 & & 
+ \frac{9}{2}\saverage{\Phi^2}\Omega_m H\left[1-\frac{2}{3}Hg_I\Omega_m\left(1-\frac{3}{2}Hg_I\right)\right]  + \frac{2(1+z)^2}{27\Omega^2H^3}\left[\saverage{\partial_k\dot\Phi\partial^k\dot\Phi}+2H \left(1+\frac{9}{2}\Omega_m\right)\saverage{\partial_k\dot\Phi\partial^k\Phi}\right.\nonumber\\
 & & 
+ \left. H^2(1+9\Omega_m)\saverage{\partial_k\Phi\partial^k\Phi}\right] + \frac{2(1+z)^2}{3\Omega_m H^2}\left[\left(1-\frac{2}{3}H g_I\right)\saverage{\Phi\partial^2\dot\Phi }+\left(1-\frac{2}{3}Hg_I\right)H\saverage{\Phi\partial^2\Phi} \right.\nonumber\\
 & & 
- \left. \left(1-\frac{2}{3}H g_I\right) \saverage{\partial^2\dot \Phi}\saverage{\Phi}-\left(1-\frac{2}{3}H g_I\right) H \saverage{\partial^2 \Phi}\saverage{\Phi}-\Omega_mH^2g_I\saverage{\Phi\partial^2\Phi}\right]+ \frac{3}{2}\saverage{A_2}H^2\Omega_m g_I\mbox{ ,}
  \end{eqnarray}
 where $ g_I= \frac{1}{g(t)}\int_0^t g(t')dt'$.  The analytical computation in this paper was performed with the help of  xAct \cite{Brizuela:2008ra}.
\end{widetext}

\bibliography{cosmoref} 

\end{document}

%% file: def2.tex
\def\3nab{\tilde{\nabla}}

\def\hsp5{\hspace{5mm}}

\def\case#1/#2{\textstyle\frac{#1}{#2}}

\def\be {\begin{equation}}
\def\ee {\end{equation}}
\def\ber {\begin{eqnarray}}
\def\eer {\end{eqnarray}}
\def\bea {\begin{eqnarray}}
\def\eea {\end{eqnarray}}

\def\bc {\begin{center}}
\def\ec {\end{center}}
\def\case#1/#2{\frac{#1}{#2}}

\newcommand{\sr}{{\cal R}}
\newcommand{\CR}{{\cal R}}

\newcommand{\CD}{{\cal D}}

\newcommand{\saverage}[1]{\left\langle #1 \right\rangle}
\newcommand{\average}[1]{\left\langle #1 \right\rangle_{\CD}}

\def\beq{\begin{equation}}
\def\eeq{\end{equation}}
\def\bea{\begin{eqnarray}}
\def\eea{\end{eqnarray}}
\def\ba{\begin{align}}
\def\ea{\end{align}}
\def\d{{\rm d}}